\documentclass[nonacm,sigconf]{acmart}

\usepackage{amsmath,amsfonts}
\usepackage{amsthm}
\usepackage{algorithm}
\usepackage{cleveref}
\usepackage{listings}
\usepackage{booktabs}
\usepackage{titletoc}
\usepackage[english]{babel}
\usepackage{subcaption}
\usepackage{balance}
\usepackage{lscape}
\usepackage{framed}

\crefname{lstlisting}{Listing}{Listings}

\newtheorem{definition}{Definition}[section]

\definecolor{findingbg}{gray}{0.95}
\definecolor{findingframecolor}{gray}{0.5}
\definecolor{sqlbg}{RGB}{245, 245, 250}
\definecolor{sqlframe}{RGB}{70, 130, 180}
\definecolor{jsonbg}{RGB}{250, 248, 245}
\definecolor{jsonframe}{RGB}{139, 90, 43}
\definecolor{codegray}{RGB}{100, 100, 100}

\newenvironment{findingbox}{%
  \MakeFramed{\advance\hsize-2\fboxsep \FrameRestore}%
}{\endMakeFramed}

\lstdefinestyle{sqlstyle}{
    language=SQL,
    backgroundcolor=\color{sqlbg},
    basicstyle=\ttfamily\small,
    keywordstyle=\color{sqlframe}\bfseries,
    commentstyle=\color{codegray}\itshape,
    showstringspaces=false,
    breaklines=true,
    breakatwhitespace=true,
    columns=fullflexible,
    frame=single,
    rulecolor=\color{sqlframe},
    framerule=1pt,
    xleftmargin=6pt,
    xrightmargin=6pt,
    framexleftmargin=6pt,
    framexrightmargin=6pt,
    framextopmargin=4pt,
    framexbottommargin=4pt,
    morekeywords={date_histogram, concat, LIMIT, per, bucket}
}
\lstnewenvironment{sqlbox}[2][]{%
  \lstset{style=sqlstyle, title=\textbf{#2}, #1}%
}{}

\lstdefinestyle{jsonstyle}{
    backgroundcolor=\color{jsonbg},
    basicstyle=\ttfamily\small,
    showstringspaces=false,
    breaklines=true,
    columns=fullflexible,
    frame=single,
    rulecolor=\color{jsonframe},
    framerule=1pt,
    xleftmargin=6pt,
    xrightmargin=6pt,
    framexleftmargin=6pt,
    framexrightmargin=6pt,
    framextopmargin=4pt,
    framexbottommargin=4pt,
    literate=
        *{:}{{{\color{jsonframe}:}}}1
        {,}{{{\color{jsonframe},}}}1
        {\{}{{{\color{jsonframe}\{}}}1
        {\}}{{{\color{jsonframe}\}}}}1
        {[}{{{\color{jsonframe}[}}}1
        {]}{{{\color{jsonframe}]}}}1
}
\lstnewenvironment{jsonbox}[2][]{%
  \lstset{style=jsonstyle, title=\textbf{#2}, #1}%
}{}

\lstdefinestyle{jsonstylecompact}{
    backgroundcolor=\color{jsonbg},
    basicstyle=\ttfamily\footnotesize,
    showstringspaces=false,
    breaklines=true,
    columns=fullflexible,
    frame=single,
    rulecolor=\color{jsonframe},
    framerule=0.5pt,
    xleftmargin=4pt,
    xrightmargin=4pt,
    framexleftmargin=4pt,
    framexrightmargin=4pt,
    framextopmargin=2pt,
    framexbottommargin=2pt,
    aboveskip=6pt,
    belowskip=6pt
}
\lstnewenvironment{jsonboxcompact}[2][]{%
  \lstset{style=jsonstylecompact, title=\textbf{\small #2}, #1}%
}{}

\newcommand{\finding}[2]{%
  \begin{findingbox}
  \textbf{Finding #1:} #2
  \end{findingbox}
}

\lstdefinestyle{agentreport}{
  basicstyle=\ttfamily\small,
  backgroundcolor=\color{gray!10},
  frame=single,
  rulecolor=\color{gray!50},
  breaklines=true,
  breakatwhitespace=false,
  columns=fullflexible,
  keepspaces=true,
  showstringspaces=false,
  tabsize=2,
  xleftmargin=1em,
  framexleftmargin=0.5em,
  aboveskip=1em,
  belowskip=1em,
  morekeywords={[1]You, Network, Security, Data, Answer, FINAL, ANSWER, CITED, CHUNKS},
  keywordstyle={[1]\bfseries\color{blue!70!black}},
  morekeywords={[2]EXACTLY, Not, found},
  keywordstyle={[2]\bfseries\color{red!70!black}},
  morecomment=[l]{\#},
  commentstyle=\bfseries\color{green!40!black},
}

\AtBeginDocument{%
  }

\begin{document}

\title{Retrieval-Augmented LLMs for Security Incident Analysis}
\author{Xavier Cadet}
\orcid{0000-0002-8545-0371}
\email{xfc@dartmouth.edu}
\affiliation{%
  \institution{Dartmouth College}
  \city{Hanover}
  \state{NH}
  \country{USA}
}

\author{Aditya Vikram Singh}
\orcid{0009-0009-8365-0503}
\affiliation{%
  \institution{Northeastern University}
  \city{Boston}
  \state{MA}
  \country{USA}
}

\author{Harsh Vijay Mamania}
\orcid{0009-0001-4680-5686}
\affiliation{%
  \institution{Northeastern University}
  \city{Boston}
  \state{MA}
  \country{USA}
}

\author{Edward Koh}
\orcid{0009-0004-2049-6274}
\affiliation{%
  \institution{Dartmouth College}
  \city{Hanover}
  \state{NH}
  \country{USA}
}

\author{Alex Fitts}
\orcid{0000-0002-8928-6011}
\affiliation{%
  \institution{PUNCH Cyber Analytics}
 \city{Reston}
 \state{VA}
 \country{USA}
}

\author{Dirk Van Bruggen}
\orcid{0009-0000-4388-9653}
\affiliation{%
  \institution{PUNCH Cyber Analytics}
  \city{Reston}
  \state{VA}
  \country{USA}
}

\author{Simona Boboila}
\orcid{0009-0003-3411-8912}
\affiliation{%
  \institution{Northeastern University}
 \city{Boston}
 \state{MA}
  \country{USA}
}

\author{Peter Chin}
\orcid{0000-0002-1913-4223}
\affiliation{%
  \institution{Dartmouth College}
  \city{Hanover}
  \state{NH}
  \country{USA}
}

\author{Alina Oprea}
\orcid{0000-0002-4979-5292}
\affiliation{%
  \institution{Northeastern University}
  \city{Boston}
  \state{MA}
  \country{USA}
}
\renewcommand{\shortauthors}{Cadet et al.}

\begin{abstract}
Investigating cybersecurity incidents requires collecting and analyzing evidence from multiple log sources, including intrusion detection alerts, network traffic records, and authentication events. This process is labor-intensive: analysts must sift through large volumes of data to identify relevant indicators and piece together what happened. We present a RAG-based system that performs security incident analysis through targeted query-based filtering and LLM semantic reasoning. The system uses a query library with associated MITRE ATT\&CK techniques to extract indicators from raw logs, then retrieves relevant context to answer forensic questions and reconstruct attack sequences. We evaluate the system with eight LLM configurations on malware traffic incidents and a multi-stage Active Directory attack. We find that LLMs have different performance and tradeoffs, with Claude Sonnet~4 achieving 94\% and DeepSeek~V3 achieving 89\% average recall across 17 malware scenarios, while DeepSeek costs 15$\times$ less than Claude per analysis, and locally-deployed Llama~3.1:70b achieves 81\% recall at zero per-query cost. Attack step detection on the Active Directory scenario reaches 100\% precision and up to 96\% recall with an enumeration prompt. These results demonstrate that combining targeted query-based filtering with RAG-based retrieval---confirmed essential by ablation studies---enables accurate, cost-effective security analysis within LLM context limits.
\end{abstract}



\maketitle

\section{Introduction}
\label{sec:introduction}

Security Operations Centers operate under increasing pressure: cyber threats grow in both volume and complexity, yet the analyst workforce has not kept pace. Modern Security Information and Event Management (SIEM) deployments routinely generate thousands of alerts per day, but empirical studies show analysts can only examine a small subset, resulting in fatigue and overlooked incidents~\cite{singhLLMsSOCEmpirical2025,kerstenGiveMeStructure2023,tariqAlertFatigueSecurity2025}. Investigating a security incident (identifying compromised assets, tracing attacker actions, understanding intrusion mechanisms) demands synthesis of evidence spread across intrusion detection systems, network telemetry, and endpoint logs. This labor-intensive work strains limited analyst capacity, motivating research into intelligent systems that can accelerate incident analysis.

Current detection mechanisms depend heavily on signatures and static rules, which cannot generalize to previously unseen attack patterns. Sophisticated adversaries exploit this gap through techniques like misconfigured certificate template abuse (e.g., ESC1~\cite{schroeder2021certified}), Kerberos ticket forgery~\cite{mitre_t1558}, and stepwise lateral movement through network trust relationships. Such attacks can progress over hours or days, depositing forensic artifacts across disparate log sources (Suricata alerts, Zeek connection records, Windows authentication events) where no individual entry reveals the complete attack chain. Recent machine learning approaches address alert triage and anomaly scoring~\cite{songAuditLLMMultiAgentCollaboration2024, weiCORTEXCollaborativeLLM2025}, but these methods yield classifications or rankings, not the narrative reconstruction that incident responders require.

Large Language Models (LLM) are well-positioned to address this problem: they can interpret unstructured text, integrate scattered facts into coherent narratives, and communicate findings in natural language~\cite{xuLargeLanguageModels2025}. Feeding raw security logs directly into an LLM, however, is impractical. A single incident may produce millions of log entries, vastly exceeding model context limits~\cite{caiLLMLogAnalyzerClusteringBasedLog2025}. These logs mix genuine attack indicators with overwhelming volumes of routine activity (normal authentications, benign connections, spurious alerts), burying the attack signal. Additionally, incident response often demands rapid turnaround, while LLM inference remains comparatively slow. These challenges motivate the design of new AI-enabled security systems that extract relevant events from such large volumes of logs and generate incident reports interpretable by security analysts.

We present a Retrieval-Augmented Generation (RAG)-based system for security incident analysis that extracts relevant threat indicators by running targeted queries on security logs, mapping the attack indicators to the MITRE ATT\&CK framework, and feeding the subsequent results to an LLM with context augmented by semantically similar events to generate interpretable reports of security incidents. The framework operates in two modes: forensic question-answering and attack reconstruction. In the first mode, analysts pose specific questions (identifying infected hosts, command-and-control (C2) servers, or initial compromise vectors) and receive evidence-backed reports with explainable answers. In the second mode, the system reconstructs attack sequences by associating distinct forensic events into causal narratives of attack progression over time. Central to our approach is a security-aware context construction pipeline. Rather than passing raw logs to the LLM, we extract through targeted queries Indicators Of Compromise (IOCs), compute aggregates of log data, and organize results into semantically meaningful chunks that preserve cross-event relationships. The RAG architecture is essential in extracting the most relevant information from multiple security logs for inclusion in the LLM context, enabling the model to generate interpretable incident reports for security analysts.
Our general architecture supports multiple LLM models---both cloud APIs and locally deployed models---enabling organizations to balance performance, cost, and data sensitivity.

We make the following contributions:

\textbf{Problem formulation for LLM-based incident analysis.} We frame security incident analysis as two complementary tasks: (a) structured question-answering, where the system answers forensic questions about indicators, hosts, and attacker actions using retrieval-augmented context, and (b) attack sequence reconstruction that correlates discrete forensic events into causal narratives of attack progression, evaluated against reference timelines.

\textbf{Security-aware RAG architecture with cross-event semantic reasoning.} We design a retrieval pipeline tailored to security data: targeted queries extract indicators, aggregators compress thousands of events into patterns, and semantic chunking preserves cross-event relationships. This enables LLMs to identify attacks requiring correlation across discrete events, such as the detection of Active Directory certificate abuse by recognizing Subject/SAN mismatches across certificate request and issuance operations. This demonstrates contextual interpretation that distinguishes coordinated exploitation from individually legitimate activities.

\textbf{LLM model evaluation with cost-effectiveness analysis.} We evaluate eight LLM configurations spanning cloud APIs (Claude Sonnet 4, DeepSeek V3, GPT-4, GPT-4o, GPT-5-mini, GPT-5.2) and local deployment (Llama 3.1:70b, Cisco Foundation-Sec-8B) on malware traffic and Active Directory attack scenarios. The best-performing configurations (Claude Sonnet 4, DeepSeek V3) achieve 94\% and 89\% average recall respectively across 17 malware scenarios, while attack step detection on Active Directory scenarios reaches 100\% precision and up to 96\% recall with an enumeration prompt. DeepSeek achieves near-equivalent accuracy to Claude at 15$\times$ lower cost, making it the most cost-effective cloud model. Local deployment (Llama~3.1:70b) provides data privacy with acceptable accuracy (81\% recall) for environments with strict data governance requirements. Notably, the security-specialized Cisco Foundation-Sec-8B model does not outperform general-purpose models at 71\% recall, suggesting that reasoning capability matters more than domain-specific pretraining for complex forensic tasks. The system completes malware analysis in 1--2 minutes per incident with cloud models, demonstrating real-time feasibility for incident response. Beyond aggregate recall, we analyze cross-source reasoning using structural recall and corroboration behavior metrics, showing that cross-chunk synthesis ability---not deployment modality or domain specialization---is the primary driver of performance differences across model families.

\textbf{Baseline comparisons and failure mode analysis.} We compare RAG against no-RAG and Suricata-only baselines, showing that preprocessing and access to multiple log sources are both necessary: no-RAG misses all attack infrastructure due to context limits, while Suricata-only misses identity attribution because IDS data lacks authentication records. We characterize three disjoint failure mode categories (query coverage, retrieval and size limits, LLM reasoning errors), each with concrete remediation strategies, providing actionable guidance for practitioners.

To support reproducibility, we release our implementation, including the query library, RAG pipeline, evaluation scripts, and pre-computed LLM reports for all 17 malware scenarios.\footnote{\url{https://github.com/neu-nds2/llm-sec-incident-analysis}} The malware traffic datasets are publicly available~\cite{malware_traffic_exercises, unit42_wireshark, sans_isc_forensics}.

\section{Related Work}
\label{sec:related_work}

\textbf{LLMs for Cybersecurity.} Large language models have emerged as tools for cybersecurity automation, with recent surveys cataloging applications across vulnerability detection, malware analysis, threat intelligence, and incident response~\cite{xuLargeLanguageModels2025}. This body of work demonstrates both the potential and limitations of LLMs for security tasks.

Early efforts adapted pre-trained models to cybersecurity text. SecureBERT~\cite{aghaeiSecureBERTDomainSpecificLanguage2023} introduced a domain-specific BERT model trained on cybersecurity corpora, demonstrating improved performance on tasks involving MITRE ATT\&CK terminology. CySecBERT~\cite{bayerCySecBERTDomainAdaptedLanguage2024} applied domain-adaptive pre-training to a large-scale cybersecurity corpus to improve performance on cybersecurity NLP tasks such as Named Entity Recognition and Cyber Threat Intelligence. For threat taxonomy mapping, TRAM fine-tunes SciBERT to map reports to MITRE ATT\&CK techniques~\cite{mitre2023threat}. These models provide strong semantic representations for classification and embedding tasks.

Log-based anomaly detection represents a key application area. LogLLM~\cite{guanLogLLMLogbasedAnomaly2025} combines BERT for semantic extraction with Llama for sequence classification on system log datasets. LogPrompt~\cite{liuLogPromptPromptEngineering2024} explores zero-shot anomaly detection through prompt engineering. These approaches focus on system logs (e.g., HDFS, Linux) and binary anomaly labels rather than security-specific logs with multi-stage attack patterns.

Recent work explores Multi-agent LLM systems to cyber security tasks such as insider threat detection (Audit-LLM~\cite{songAuditLLMMultiAgentCollaboration2024} demonstrates reduced false positives) while SOC-specific systems like  CORTEX~\cite{weiCORTEXCollaborativeLLM2025} propose specialized agents for alert triage. Work on CTI automation~\cite{tsengUsingLLMsAutomate2024} develops agents that extract Indicators of Compromise from threat reports. These systems advance alert-level processing but focus on triage and summarization rather than answering specific questions about detected incidents. 
Examining operational use, Singh et al. analyzed queries from 45 SOC analysts, finding that LLMs function as cognitive aids that augment rather than replace expertise~\cite{singhLLMsSOCEmpirical2025}.

In contrast to prior work focusing on alert triage or classification, our work addresses end-to-end incident analysis formulated as question-answering over heterogeneous security logs, processing data from intrusion detection systems, network monitors, and authentication events.

\textbf{Retrieval-Augmented Generation.} Retrieval-Augmented Generation (RAG) addresses limitations of LLMs—hallucination, outdated knowledge, and opacity—by conditioning generation on external evidence retrieved at inference time~\cite{lewisRetrievalAugmentedGenerationKnowledgeIntensive2020}. RAG has become the dominant paradigm for knowledge-intensive tasks where accuracy and traceability are required. Surveys categorize RAG architectures into retriever-centric, generator-centric, and hybrid designs, with recent advances in adaptive retrieval, iterative refinement, and self-reflection~\cite{gaoRetrievalAugmentedGenerationLarge2024, sharma2025ragsurvey}.

Recent work applies RAG to security tasks. CyberRAG~\cite{blefariCyberRAGAgenticRAG2026} presents an agentic framework for cyber attack classification, using iterative retrieval over domain-specific knowledge bases to reduce false positives and generate explanations. AgCyRAG~\cite{kurniawanAgCyRAGAgenticKnowledge2025} combines knowledge graph retrieval with vector search for security log analysis, orchestrating specialized agents to integrate structured threat intelligence with security log data. Work on threat intelligence automation~\cite{tsengUsingLLMsAutomate2024} uses RAG to extract Indicators of Compromise from CTI reports and generate regular expressions to assist security analysts in establishing SIEM correlation rules.
GraphRAG-based approaches~\cite{carvalhoApplicationGraphRAGNetwork2025} leverage ontology-backed knowledge graphs constructed from network logs to support LLM-driven query generation for network security monitoring.
These systems demonstrate RAG's applicability to security data, primarily focusing on attack classification and threat intelligence extraction.

Our work applies RAG to a different task: answering structured security questions over heterogeneous log sources (Suricata alerts, Zeek connections, Windows events). This requires specialized chunking strategies tailored to security data
and retrieval that supports reasoning about attack progression across multiple log sources.

\section{Problem Formulation}
\label{sec:formulation}

We formalize security incident analysis as a question-answering task and define the evaluation metrics used throughout our experiments.

\subsection{Security Incident Analysis as Question Answering}

Let $I \in \mathcal{I}$ denote a \textbf{security incident}, characterized by a collection of security logs $L_I$ (e.g., Suricata alerts, Zeek connection logs, Windows event logs). We formulate incident analysis as question-answering: given an incident's logs, the system must answer forensic questions about the attack.

\begin{definition}[Evaluation Questionnaire]
A \textbf{questionnaire} $Q \sim \mathcal{Q}(I)$ is a finite collection of
question-answer pairs drawn from the space of possible questionnaires over
incident $I$:
$$Q = \{(q_i, a_i)\}_{i=1}^{M}$$
where:
\begin{itemize}
    \item $q_i \in \mathcal{Q}$ is a natural language question
    \item $a_i$ is the reference answer
    \item $M$ is the number of questions in the questionnaire
\end{itemize}
\end{definition}

Beyond forensic identification, we also evaluate attack reconstruction: deducing the attack path a malicious actor took over a given timeframe. For this task, we pose a compound question (``What did the attacker do? What defenses should be implemented?'') and evaluate both attack steps and defense recommendations using precision (fraction of reported items matching the reference) and recall (fraction of reference items correctly identified). This formulation unifies both modes within the question-answering framework, although attack reconstruction requires richer evaluation against known attack timelines (Section~\ref{sec:ad-attack}).

\subsection{Security Analyzer}
\label{subsec:analyzer_def}

The Security Analyzer takes raw security logs as input and produces answers to security questions. We decompose it into two modular components that can be tuned and improved independently.

\begin{definition}[Security Analyzer]
A \textbf{Security Analyzer} $SA$ is a function:
$$SA : \mathcal{L} \times \mathcal{Q} \rightarrow \mathcal{R}$$
where $\mathcal{L}$ is the space of security logs, $\mathcal{Q}$ is the question
space, and $\mathcal{R}$ is the response space. For an incident $I$ with logs
$L_I \in \mathcal{L}$ and questionnaire $Q = \{(q_i, a_i)\}_{i=1}^{M}$:
$$SA(L_I, Q) = \{r_i\}_{i=1}^{M} \quad \text{where } r_i = SA(L_I, q_i)$$
\end{definition}

\paragraph{Security Context Extraction} ($SCE : \mathcal{L} \times \mathcal{Q} \rightarrow \mathcal{C}$): This component interfaces with the SIEM platform to retrieve logs and applies a library of predefined IOC queries $\mathcal{F} = \{f_1, \ldots, f_K\}$ to extract security-relevant features such as Suricata alerts, suspicious certificates, high-volume connections, authentication anomalies. A RAG retrieval function then selects the top-k most relevant chunks for each question based on semantic similarity.

\paragraph{RAG-LLM Analysis} ($LLM : \mathcal{C} \times \mathcal{Q} \rightarrow \mathcal{R}$): Given the retrieved context and a question, the LLM produces a structured response with supporting evidence.
This modularity allows each component to be tuned and improved independently: the query set $F$ determines detection coverage, the embedding model controls retrieval quality, $k$ governs context size, and the LLM affects reasoning accuracy.

\section{System Architecture}
\label{sec:architecture}

\begin{figure*}[t]
\centering
\includegraphics[width=\textwidth]{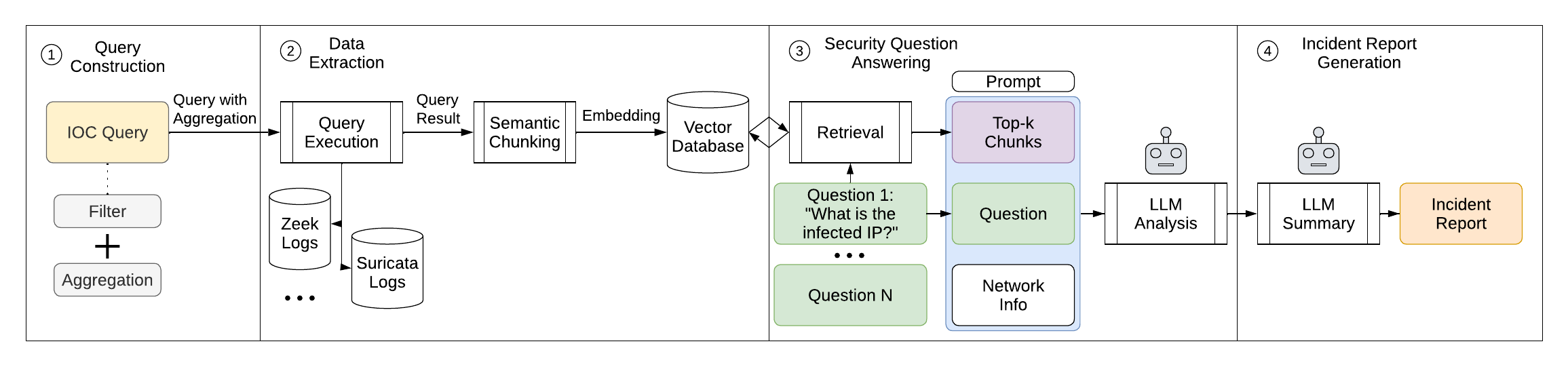}
\caption{System architecture. Security Context Extraction queries the SIEM data, filtering the traffic to expose indicators of compromise. RAG-LLM Analysis embeds this data, retrieves relevant context for each question, and prompts an LLM to generate an incident report.}
\Description{System architecture. Security Context Extraction queries the SIEM data, filtering the traffic to expose indicators of compromise. RAG-LLM Analysis embeds this data, retrieves relevant context for each question, and prompts an LLM to generate an incident report.}
\label{fig:architecture}
\end{figure*}

Our architecture comprises two components: Security Context Extraction, which reduces raw logs to security-relevant information through targeted queries, and RAG-LLM Analysis, which retrieves relevant context and prompts an LLM to answer forensic questions. Figure~\ref{fig:architecture} illustrates the overall architecture.

\subsection{Design Principles}
\label{subsec:design_principles}

Five principles guided our architectural decisions.

\textbf{1) Filter before analysis.} 
LLMs are too slow to process large volumes of raw logs and perform poorly on noisy data. We first extract security-relevant information using targeted IOC queries, allowing the system to scale to tractable log volumes and generate interpretable reports. 

\textbf{2) Embed once, query many times.}
Extracted data is semantically embedded into a vector database and cached to disk. The index is built once per dataset and can be queried repeatedly with different questions, LLMs, or prompt strategies without reprocessing.

\textbf{3) Periodic analysis.} 
Rather than invoking the LLM on every event, the system performs analysis at defined intervals, batching related activity into coherent time windows.

\textbf{4) Provider-agnostic design.} 
LLM interactions are abstracted behind a common interface, enabling substitution of providers to balance cost, latency, and privacy requirements.

\textbf{5) Extensible IOC query library.} 
Detection queries are maintained as a separate library with associated MITRE ATT\&CK techniques, allowing extension without modifying the core pipeline.

\subsection{Security Context Extraction}
\label{sec:extraction}

The system integrates with existing network monitoring infrastructure through standard APIs. Our implementation uses Security Onion, an open-source platform that aggregates network traffic using Suricata and Zeek, communicating through ElasticSearch REST API calls. The architecture generalizes to any SIEM exposing a query API.

Security Context Extraction relies on a library of IOC extraction queries, each targeting a specific activity type. The AD attack library covers 9 query types mapping to MITRE ATT\&CK techniques (certificate abuse, Kerberos ticket forgery, lateral movement, etc.); malware traffic scenarios use a separate set targeting network-level IOCs. The full query-to-technique mappings and scoping notes are given in \Cref{app:mitre-coverage}; full query definition examples appear in Appendix~\ref{appendix:aggregation-queries}.

The current query library was developed iteratively for the evaluated attack types, validating each query against known attack data; extending coverage to novel attack patterns requires adding corresponding queries, though the modular design facilitates this.

Each query pairs a base filter with aggregators that compress matching events into ranked patterns, exposing behavioral signals that individual log entries would not reveal. This design is an architectural choice: the aggregation step reduces thousands of events into compact, semantically rich summaries that preserve cross-event patterns required for attack identification. The extracted aggregations are then embedded and indexed for retrieval, as described in the following section.

\subsection{RAG-LLM Analysis}
\label{sec:analyzer}

The RAG-LLM Analysis component transforms extracted security information into actionable incident reports. Given a set of security questions, it retrieves relevant context from the embedded document store and prompts an LLM to produce evidence-based findings. This design decouples index construction from analysis, enabling rapid experimentation with different questions, providers, and prompt strategies without reprocessing the source data.

\textbf{Data Processing and Embedding.}
The Security Context Extraction component produces structured output containing aggregations and log entries from the SIEM platform. Before embedding, this output is split into semantically meaningful chunks rather than arbitrary text segments. The data chunks consist primarily of aggregate results, which summarize patterns across many events. 

Chunks are embedded using a pre-trained sentence embedding model and indexed in a vector store for fast similarity search.\footnote{Our implementation uses all-mpnet-base-v2 and FAISS; other embedding models and vector stores could be substituted.} The complete index—chunks, metadata, and embeddings—is serialized to disk, allowing the embedding phase to run once per dataset.

\textbf{Retrieval and Prompt Construction.}
Given a security question, the system retrieves relevant chunks from the embedding index using semantic similarity search. The question is embedded using the same model as the indexed chunks, and nearest neighbors are retrieved based on vector similarity. 
This retrieval-augmented approach grounds LLM responses in actual log data, reducing hallucination and enabling evidence-based reasoning that analysts can verify.

The prompt establishes the LLM's role as a cybersecurity analyst, presents the question, and provides brief network context with topology information (address ranges, domain name, key infrastructure). Retrieved chunks appear next, each tagged with its source file to support attribution. The prompt concludes with an instruction to answer based on the provided evidence or indicate when information is insufficient. We provide the prompt template in Appendix~\ref{app:base_prompt} and a sample network context in Appendix~\ref{app:network_context}.

\textbf{Security Questions.}
The system supports different categories of security questions depending on the analysis goal (Appendix~\ref{app:questions}). \textit{Identification questions} focus on indicators of compromise: infected hosts, compromised accounts, malicious domains, and command-and-control servers. \textit{Timeline questions} focus on attack progression: what steps did the attacker take, in what order, and when. \textit{Remediation questions} focus on defensive actions: what steps should be taken to contain the incident and prevent further damage. All question types use the same retrieval and prompting pipeline, with the LLM adapting its response format based on the question framing. 

\textbf{Multi-LLM Support.}
Following design principle number 4, the system abstracts LLM interactions behind a common interface, allowing different providers to be substituted without modifying the analysis pipeline. Currently supported providers include Anthropic, OpenAI, and DeepSeek as cloud APIs, and Ollama (Llama~3.1:70b) and Cisco as locally-deployed options. 

\textbf{Output Generation.}
The system produces a structured incident report containing findings for each security question, with responses that include specific evidence (IP addresses, hostnames, timestamps) and attribute findings to source files (Appendix~\ref{app:anthropic_report}). A summary section follows, condensing the findings into a brief overview: attack type, key indicators, and recommended actions. The report concludes with metadata for reproducibility: the LLM provider, the number of chunks analyzed, the retrieval settings, and the timing information.

\subsection{End-to-End Example}
\label{sec:example}

To illustrate how extraction and RAG-LLM analysis work together, consider a malware infection where the compromised host must be identified along with the responsible user account. The raw dataset contains 3,694 log events spanning HTTP connections, Kerberos authentication, IDS alerts, and TLS certificate exchanges (Fake Authenticator Malware scenario~\cite{malware_traffic_exercises}).

The extraction queries reduce this to semantically rich summaries. The Suricata alert query identifies malicious activity:

\begin{jsonboxcompact}{Suricata High-Severity Alert Aggregation}
{"high_severity_signatures": [
  {"key": "ET MALWARE Fake Microsoft Teams CnC", "doc_count": 2},
  {"key": "ET INFO PS1 Powershell File Request", "doc_count": 2}],
 "high_severity_sources": [{"key":"10.1.17.215", "doc_count": 6}]}\end{jsonboxcompact}

These alerts confirm malicious activity from IP \texttt{10.1.17.215}, but reveal neither the user responsible nor the specific payloads retrieved. The Kerberos client query provides identity attribution:

\begin{jsonboxcompact}{Kerberos Client Aggregation}
{"key": "shutchenson", "doc_count": 11, 
 "source_ips": [{"key": "10.1.17.215", "doc_count": 11}]}
{"key": "DESKTOP-L8C5GSJ", "doc_count": 10,
 "source_ips": [{"key": "10.1.17.215", "doc_count": 10}]}
\end{jsonboxcompact}

The file download query reveals the attack payload:

\begin{jsonboxcompact}{File Download Aggregation}
{"key": "10.1.17.215 -> 5.252.153.241:80 : /api/file/get-file/29842.ps1",
 "doc_count": 4}
\end{jsonboxcompact}

No single data source provides the complete picture. The IDS alerts lack user attribution; Kerberos logs show authentication but not malicious behavior; HTTP logs capture downloads but cannot distinguish malicious from benign traffic. When these aggregations are embedded and retrieved, the LLM corroborates across sources via the shared IP address, concluding that user \texttt{shutchenson} on workstation \texttt{DESKTOP-L8C5GSJ} downloaded a PowerShell payload from a command-and-control server.

This example illustrates how extraction reduces thousands of log events into three compact aggregations while preserving the semantic links (shared IP addresses) that enable cross-source reasoning. The complete queries and corresponding LLM output appear in Appendix~\ref{appendix:aggregation-queries}.

\section{Experimental Overview}
\label{sec:experiments}

We evaluate the system on two complementary settings. The first is malware traffic analysis: 17 scenarios drawn from Malware-Traffic-Analysis.net, Unit~42 Wireshark Quizzes, and SANS~ISC contests, evaluated as forensic question-answering (\Cref{sec:malware}).  The second is an Active Directory attack reconstruction with associated defense recommendation (\Cref{sec:ad-attack}).

\textbf{Research Questions.} We explore the following questions:
\begin{description}
    \item[RQ1 (Feasibility)] Can a RAG-based system achieve acceptable performance on security incident analysis?
    \item[RQ2 (Model Comparison)] How do different LLM configurations compare in performance, latency, and cost?
    \item[RQ3 (Context Sensitivity)] How does the number of retrieved chunks $k$ affect performance?
    \item[RQ4 (RAG Necessity)] Is RAG preprocessing essential, or can LLMs analyze raw security logs directly?
    \item[RQ5 (Practical Utility)] Do system-generated defense recommendations align with expert-defined actions?
\end{description}

RQ1--RQ4 are addressed in the malware traffic evaluation (Section~\ref{sec:malware}), which compares eight LLM configurations across 17 scenarios and evaluates a no-RAG baseline. RQ1 and RQ5 are addressed in the Active Directory evaluation (Section~\ref{sec:ad-attack}), which tests attack detection and defense recommendation on a multi-stage intrusion.

\textbf{LLM Configurations.}
We evaluate eight configurations spanning cloud APIs and local deployment:

\begin{itemize}
     \item \textbf{Claude Sonnet 4} (Anthropic): Cloud API
     \item \textbf{DeepSeek V3} (DeepSeek): Cloud API
     \item \textbf{GPT-4, GPT-4o, GPT-5-mini, GPT-5.2} (OpenAI): Cloud API
     \item \textbf{Llama 3.1:70b} (via Ollama): Local GPU deployment
    \item \textbf{Cisco Foundation-Sec-8B}: Local GPU deployment, security-specialized
\end{itemize}

\textbf{RAG Configuration.}
The RAG pipeline uses the following parameters:
\begin{itemize}
    \item \textbf{Embedding model:} all-mpnet-base-v2 (sentence transformer)
    \item \textbf{Vector store:} FAISS with cosine similarity
    \item \textbf{Retrieved chunks:} $k=7$ per question (default); Section~\ref{sec:context-size} examines sensitivity to $k$
    \item \textbf{LLM parameters:} temperature$=0.1$, max output tokens$=4000$
\end{itemize}

Each chunk contains a complete aggregation result rather than arbitrary text segments, preserving semantic coherence for retrieval.

\textbf{Evaluation Metrics.}
We evaluate against reference sets derived from published scenario solutions and manual log analysis. Let $R$ denote the reference set and $A$ denote the model's answer set. We compute:
$$\text{Precision} = \frac{|A \cap R|}{|A|}, \quad \text{Recall} = \frac{|A \cap R|}{|R|}$$
For single-value questions (e.g., infected host IP), precision equals recall and scoring is binary. For set-valued questions (e.g., C2 indicators), models receive partial credit proportional to the fraction of reference indicators found.

The malware evaluation primarily reports recall: reference answers capture \emph{confirmed} malicious indicators from published solutions authored and reviewed by domain experts, not an exhaustive enumeration. Models may identify additional legitimately suspicious indicators absent from the reference set; Appendix~\ref{appendix:precision} provides full precision-recall breakdowns. The AD evaluation reports both precision and recall, as its reference timeline was manually extracted through exhaustive log review of the Security Onion Kibana dashboard.
\section{Malware Traffic Analysis}
\label{sec:malware}

Evaluating LLM-based security analysis is constrained by the absence of public benchmarks pairing raw traffic with vetted forensic reference answers. To address this gap, we curated scenarios from three established sources whose solutions are authored and reviewed by domain experts: Unit~42 Wireshark Quizzes~\cite{unit42_wireshark} (Palo Alto Networks' threat research team), the SANS Internet Storm Center monthly forensic challenges~\cite{sans_isc_forensics} (with published official solutions), and Malware-Traffic-Analysis.net~\cite{malware_traffic_exercises}. 

We evaluate our method on 17 scenarios (Table~\ref{tab:scenario_information}), drawn from these sources.   
Our evaluation spans 321 MB of PCAPs, 94,948 Security Onion log events, 129 forensic questions, 218 reference indicators.

\subsection{Incident Analysis Performance}

Table~\ref{tab:malware-accuracy} presents recall for each model across 17 scenarios from three independent challenge sources, addressing \textbf{RQ1} (feasibility).

\begin{table}[h]
\centering
\caption{Incident Analysis Recall (\%) by Scenario and Model}
\label{tab:malware-accuracy}
\footnotesize
\begin{tabular}{lrrrrrrrr}
\toprule
Scenario & Cl. & Deep- & Ll. & Cisco & GPT & GPT & GPT & GPT \\
& & Seek &  & 8B & 4 & 4o & 5m & 5.2 \\
\midrule
apr21isc       & 92  & 79  & 79 & 71  & 85 & 85 & 85 & 85 \\
burnincandle   & 100 & 93  & 84 & 77  & 70 & 70 & 70 & 70 \\
dec21isc       & 100 & 100 & 94 & 81  & 75 & 75 & 75 & 75 \\
dirty\_rat     & 89  & 75  & 71 & 43  & 50 & 64 & 50 & 50 \\
easy\_123      & 100 & 100 & 92 & 100 & 75 & 75 & 75 & 92 \\
fake\_auth     & 100 & 100 & 94 & 81  & 69 & 69 & 56 & 69 \\
gozi           & 77  & 88  & 72 & 75  & 72 & 70 & 70 & 70 \\
iced\_id       & 94  & 91  & 88 & 75  & 50 & 50 & 38 & 50 \\
jun21isc       & 98  & 80  & 70 & 38  & 54 & 60 & 54 & 54 \\
koi\_stealer   & 100 & 88  & 69 & 88  & 56 & 69 & 69 & 69 \\
lumma\_stealer & 100 & 100 & 93 & 86  & 71 & 71 & 71 & 71 \\
may21isc       & 86  & 86  & 79 & 79  & 79 & 79 & 64 & 64 \\
net\_rat       & 100 & 93  & 86 & 57  & 79 & 71 & 57 & 71 \\
oct21isc       & 88  & 78  & 72 & 48  & 46 & 58 & 46 & 58 \\
qakbot         & 90  & 78  & 84 & 69  & 69 & 62 & 69 & 62 \\
redline\_stlr  & 94  & 94  & 81 & 56  & 81 & 81 & 75 & 81 \\
warmcookie     & 96  & 93  & 79 & 86  & 68 & 68 & 68 & 68 \\
\midrule
\textbf{Average} & \textbf{94} & \textbf{89} & \textbf{81} & \textbf{71} & \textbf{67} & \textbf{69} & \textbf{64} & \textbf{68} \\
\bottomrule
\multicolumn{9}{l}{Cl.=Claude Sonnet 4; Ll.=Llama 3.1:70b} 
\end{tabular}%
\end{table}

Claude Sonnet 4 achieves the highest average recall (94\%), followed by DeepSeek V3 (89\%) and Llama 3.1:70b (81\%). Locally-deployed models perform comparably to several cloud-based models: Llama (81\%) and Cisco Foundation-Sec-8B (71\%) exceed or match most GPT variants (64--69\%), suggesting that organizations with data sensitivity requirements need not sacrifice accuracy.

The results reveal distinct capability differences across model families. GPT models consistently struggle with questions requiring cross-source integration, such as identifying C2 servers from behavioral aggregations (connection counts, file download volumes) correlated with Suricata alerts. Claude and DeepSeek achieve near-perfect recall on these questions, while GPTs often default to listing indicators from a single data source.

IcedID and the multi-host scenarios (oct21isc, jun21isc) prove most challenging overall. IcedID involves a multi-stage infection chain requiring correlation of redirect IPs, download URLs, and C2 domains across separate log sources (details in Appendix~\ref{appendix:icedid}). The multi-host scenarios (3 infected machines, 5+ malware families) increase the data volume and require the model to attribute indicators to specific hosts.

\finding{1}{Recall is driven by a model's cross-source reasoning ability, not deployment modality: locally-hosted models match or exceed cloud GPT variants. Multi-stage infections and multi-host scenarios prove most challenging, requiring cross-source correlation.}

\subsection{Reasoning and Cross-Source Integration}
\label{sec:reasoning-patterns}

Beyond recall, models exhibit qualitatively different reasoning strategies, most evident on questions that structurally require cross-source synthesis.
Forty-two of the 129 questions are set-valued with reference indicators distributed across two or more retrieval chunks---no single chunk contains the complete answer.
We measure cross-source integration with two complementary metrics: \emph{structural recall} (recall on the multi-chunk subset) and \emph{corroboration behavior} (the fraction of set-valued answers whose self-reported chunk citations span two or more chunks).

\begin{table}[h]
\centering
\caption{Cross-source integration on set-valued questions. }
\label{tab:cross_source}
\footnotesize
\begin{tabular}{lrr}
\toprule
LLM & Recall on set-valued & Multi-chunk \\
& questions, hard (\%) & cite rate, soft (\%) \\
\midrule
Claude Sonnet 4 & 89 & 95 \\
DeepSeek V3     & 78 & 88 \\
Llama 3.1:70b   & 56 & 76 \\
Cisco 8B        & 46 & 36 \\
GPT-4           & 52 & 38 \\
GPT-4o          & 53 & 31 \\
GPT-5-mini      & 47 & 33 \\
GPT-5.2         & 52 & 33 \\
\bottomrule
\end{tabular}
\end{table}

Claude reaches 89\% recall on the multi-chunk subset and cites two or more chunks for 95\% of set-valued answers; GPT-family models recall 47--53\% and cite multi-chunk evidence for only 31--38\%.
The context-size ablation (Section~\ref{sec:context-size}) corroborates this: going from $k{=}1$ to $k{=}3$, Claude/DeepSeek gain 33--36 points while GPTs gain only 13--18, consistent with stronger cross-chunk synthesis in the former.

The reasoning gap is also visible qualitatively. In Appendix~\ref{appendix:llm-reasoning}, we compare responses to the C2 detection question (Q5) on the Koi Stealer scenario.
DeepSeek identifies only the confirmed C2 server, explicitly rejecting domain-resolved IPs as insufficiently corroborated. Cisco produces verbose sequential reasoning, narrating its inspection of each data source before arriving at the same single correct answer as DeepSeek---but cites only one chunk rather than cross-validating, relying on the strongest single signal. GPT-5.2 lists several IPs without prioritization, including known Microsoft and Azure infrastructure as C2 candidates, treating the question as a data extraction task rather than an analytical one.

These patterns reflect a precision-coverage trade-off: DeepSeek minimizes false positives through evidence-weighted reasoning; Cisco reaches the same conservative answer through verbose sequential analysis but without cross-source synthesis; GPT-5.2's undifferentiated enumeration---including known cloud provider addresses as C2 candidates---increases analyst burden. For production deployment, evidence-weighted reasoning (distinguishing confirmed threats from suspicious activity) is preferable to exhaustive listing. Appendix~\ref{appendix:precision} provides detailed precision-recall breakdowns per scenario.

\subsection{Impact of Context Size}
\label{sec:context-size}

We investigate how the number of retrieved chunks $k$ affects performance (\textbf{RQ3}). We evaluate $k \in \{1, 3, 5, 7, 14\}$ across all 17 scenarios and 8 LLMs.

\begin{table}[htb]
\centering
\caption{Context-Size Ablation: Recall (\%) by $k$}
\label{tab:context-ablation}
\footnotesize
\begin{tabular}{lrrrrr}
\toprule
LLM & $k{=}1$ & $k{=}3$ & $k{=}5$ & $k{=}7$ & $k{=}14$ \\
\midrule
Claude Sonnet 4   & 39 & 72 & 86 & 94 & 94 \\
DeepSeek V3 & 35 & 71 & 80 & 89 & 89 \\
Llama 3.1:70b   & 40 & 68 & 78 & 81 & 78 \\
Cisco 8B    & 43 & 62 & 72 & 71 & 67 \\
GPT-4     & 46 & 60 & 64 & 67 & 66 \\
GPT-4o    & 46 & 59 & 65 & 69 & 64 \\
GPT-5-mini    & 44 & 62 & 65 & 64 & 66 \\
GPT-5.2   & 46 & 62 & 64 & 68 & 64 \\
\bottomrule
\end{tabular}%
\end{table}

At $k{=}1$, GPTs (46\%) outperform Claude (39\%) and DeepSeek (35\%): Claude tends to answer ``Not found'' when only one chunk is available, while GPTs extract answers from whatever single chunk they receive. By $k{=}3$ this reverses as Claude integrates across chunks (72\% vs 61\% for GPTs), and the gap widens to 27 points at $k{=}7$ (94\% vs 67\%). This crossover reveals a differential cross-chunk integration ability: Claude and DeepSeek gain 33--36 points going from $k{=}1$ to $k{=}3$, while GPTs gain only 13--18.

At $k{=}14$, recall plateaus or slightly declines for most LLMs (Cisco $-4$, Llama $-3$, GPT-4o $-5$). On the largest scenario (jun21isc, 3 infected hosts), $k{=}14$ exceeds GPT context limits entirely. $k{=}7$ is a sweet spot across all LLMs.

\finding{2}{LLMs differ in how much they benefit from additional context. The retrieval window $k$ interacts with model reasoning: more context helps only if the model synthesizes across chunks. $k{=}7$ is a sweet spot; $k{=}14$ provides no further benefit.}

\subsection{Baseline Comparisons}
\label{subsec:no-rag-baseline}

\textbf{No-RAG Baseline.} We evaluate a no-RAG baseline to assess whether RAG preprocessing is essential or whether LLMs can analyze raw security logs directly (\textbf{RQ4}).
For each security question, we send as many raw logs as fit within the model's context window, without any feature extraction or embedding. This represents the simplest possible approach: direct LLM analysis of unprocessed security data. While a sliding window approach processing logs in multiple batches could eventually cover all data, it would require $\sim$24 batches for this scenario's 3,694 logs (Claude's 158-log window), resulting in $24 \times 8 = 192$ API calls versus 8 for RAG.

\begin{table}[ht]
\centering
\caption{RAG vs No-RAG Baseline: Recall (\%) on the fake\_auth scenario}
\label{tab:no-rag-baseline}
\footnotesize
\begin{tabular}{lcccc}
\toprule
LLM & RAG (k=7) & No-RAG & Logs Seen by LLM \\
\midrule
Claude Sonnet 4 & 100 & 50 & 158 (4.3\%) \\
DeepSeek V3     & 100 & 50 & 104 (2.8\%) \\
GPT-4           & 69 & 38 & 22 (0.6\%) \\
GPT-4o          & 69 & 38 & 22 (0.6\%) \\
GPT-5-mini      & 56 & 38 & 22 (0.6\%) \\
GPT-5.2         & 69 & 38 & 22 (0.6\%) \\
\bottomrule
\end{tabular}%
\end{table}

Table~\ref{tab:no-rag-baseline} compares RAG and no-RAG performance on the Fake Authenticator scenario.
Context window constraints limit no-RAG approaches to 0.6--4.3\% of available logs (22--158 out of 3,694). The logs that fit are chronologically first and contain primarily benign startup activity. All models correctly identify victim information (IP, hostname, timestamp) present in these early logs but miss all attack infrastructure: malicious domains, C2 servers, and suspicious certificates appear only in later logs that exceeded context limits. GPT models additionally confuse the Windows machine hostname with the human user account, a reasoning flaw that persists in both RAG and no-RAG settings.

\textbf{Suricata-Only Baseline.} We also evaluated whether traditional IDS alerts alone suffice. Using only Suricata alerts from the Fake Authenticator scenario, an analyst could identify the infected IP and C2 infrastructure. However, hostname, user account, and initial infection domain are absent from IDS data entirely—these require Kerberos authentication logs and HTTP connection records. This yields 40\% recall, comparable to no-RAG but with a different failure mode: no-RAG fails due to context limits, while Suricata-only fails due to missing data sources.

\finding{3}{Both no-RAG and Suricata-only baselines achieve low recall but for different reasons: no-RAG misses attack infrastructure due to context limits; Suricata-only misses identity attribution due to lacking cross-source data. The performance gap from RAG demonstrates that both preprocessing and access to multiple log sources are essential.}

\subsection{Cost Analysis}
\label{subsec:cost-analysis}

We compare cost-effectiveness across models based on observed token usage. Table~\ref{tab:cost-comparison} summarizes pricing (from vendor documentation as of early 2026), analysis time, and performance.
The cost per analysis is computed as:
\begin{equation}
C_{\text{analysis}} = N_{\text{calls}} \cdot \left(
    \frac{T_{\text{in}}}{10^6} \cdot P_{\text{in}} +
    \frac{T_{\text{out}}}{10^6} \cdot P_{\text{out}}
\right)
\end{equation}
The number of LLM calls per analysis is $N_{\text{calls}} = 8$ (one per question). We use conservative estimates of input and output tokens per call: $T_{\text{in}} = 3000$ and $T_{\text{out}} = 400$. $P_{\text{in}}$, $P_{\text{out}}$ are the model's prices in USD per million tokens from Table~\ref{tab:cost-comparison}.

\begin{table}[t]
\centering
\caption{Timing vs Cost per LLM}
\label{tab:cost-comparison}
\footnotesize
\resizebox{\columnwidth}{!}{%
\begin{tabular}{lrrrrrr}
\toprule
LLM  & Recall & Avg time/ & Avg time/ & In & Out & Cost/ \\
 & (\%) & scenario & question & (\$/M) & (\$/M) & Anlys. \\
\midrule
Claude Sonnet 4  & 94 & 100s  & 13.1s & 3.00 & 15.00 & \$0.12  \\
DeepSeek V3      & 89 & 110s  & 14.4s & 0.28 & 0.42  & \$0.008 \\
GPT-4            & 67 & 18s   & 2.4s  & 3.00$^\dagger$  & 12.00$^\dagger$ & \$0.11$^\dagger$  \\
GPT-4o           & 69 & 19s   & 2.6s  & 2.50  & 10.00 & \$0.09  \\
GPT-5-mini       & 64 & 18s   & 2.4s  & 0.25  & 2.00  & \$0.01  \\
GPT-5.2          & 68 & 17s   & 2.3s  & 1.75 & 14.00 & \$0.09  \\
Llama 3.1:70b  & 81 & 200s  & 26.4s & ---  & ---   & \$0$^*$ \\
Cisco 8B     & 71 & 376s  & 49.5s & ---  & ---   & \$0$^*$ \\
\bottomrule
\multicolumn{7}{l}{$^\dagger$GPT-4.1 pricing used; GPT-4 pricing no longer published.} \\
\multicolumn{7}{l}{$^*$Local deployment; requires GPU infrastructure, cost not included.} \\
\end{tabular}%
}
\end{table}

\begin{figure}[h]
    \centering
    \includegraphics[width=\columnwidth]{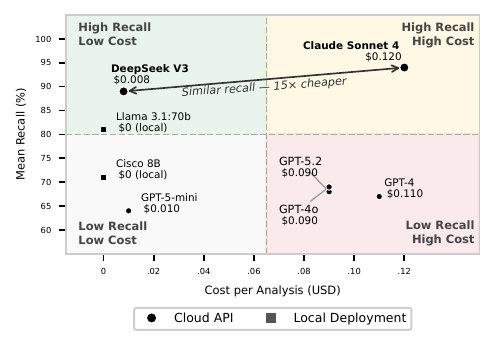}
    \caption{Cost vs.\ Recall Across LLM models.}
    \Description{Cost vs. Recall Across LLM models, showing that Claude Sonnet 4 stands in the High Recall High Cost Quadrant, most of the GPT models lie in the Low Recall High Cost  and the DeepSeek and LLama 3.1 70B land in the High Recall Low Cost quadrant.} 
    \label{fig:cost-recall}
\end{figure}

DeepSeek achieves 89\% recall at the lowest cost (\$0.008 per analysis), making it 15$\times$ cheaper than Claude while within 5 points of its recall. At operational scale (1,000 analyses/day), this translates to \$240/month versus \$3,600/month. Local deployment eliminates per-query costs but requires GPU infrastructure and incurs longer processing times: Llama (70B) achieves 81\% recall at 3.3 min/analysis, while Cisco 8B achieves 71\% at 6.3 min/analysis.

\section{Active Directory Attack Analysis}
\label{sec:ad-attack}

\textbf{Data.} The attack data originates from a controlled red team exercise against an Active Directory environment. The scenario simulates a sophisticated attacker (Red) attempting to compromise a Windows domain while defenders (Blue) respond in real-time.

\emph{Red Team Actions:} Initial compromise via Metasploit, certificate request with Administrator SAN, certificate enumeration, Kerberos TGT and service ticket requests, PSExec lateral movement to Domain Controller, malicious user creation, and ultimately SSH access to SCADA with service disruption (attack objective).

\emph{Blue Team Actions:} Certificate revocation, KRBTGT reset, Domain Controller restart, password reset, and user account disable. With defensive response, Blue successfully prevented SCADA compromise.

\textbf{Evaluation.}
Unlike the malware scenarios with discrete question-answer pairs, the AD evaluation uses the attack recreation task: given a timeframe, identify attack steps and recommend defensive actions. We evaluate on 5-minute sliding windows (15:30--15:35, 15:35--15:40, 15:40--15:46), simulating real-time analysis (window size justification in \Cref{app:window-justification}). The reference attack steps in the timeline were manually extracted through exhaustive log review via the Security Onion Kibana dashboard, enabling evaluation of both precision and recall. Each data filtering query produces a compact result fitting in a single chunk; with 9 queries generating 9 chunks, all relevant context is retrieved simultaneously. We prompt the LLM with an \emph{enumeration prompt} that instructs it to list each attack step as a separate, timestamped entry---as opposed to a narrative summary---so that individual steps can be matched against the reference timeline.

\subsection{Attack Step Detection}

We evaluate attack-step detection across 8 LLMs (\textbf{RQ1}). Recall ranges from 46\% to 96\%, with Claude Sonnet~4 and DeepSeek~V3 reaching 96\% (\Cref{tab:timestamp_recall}).
The performance gap is concentrated in the 15{:}30--35 window, where 11 distinct attack steps occur within five minutes: Claude and DeepSeek enumerate all 11, while GPT models report 2--5, leaning on Suricata alert labels and under-representing certificate, Kerberos, and service-installation evidence. The single missed step at 96\% is a query-coverage gap in the \texttt{kerberos\_auth} filter (see \Cref{sec:failure-modes}). Precision is 100\% throughout---every reported attack step corresponded to actual malicious activity. The system correctly identified initial access (Metasploit, PowerShell injection, file drops), the credential theft chain (certificate enumeration, request, issuance, Kerberos tickets), and lateral movement (PSExec via service installation). Notably, recall ordering across LLMs on the AD scenario mirrors the malware evaluation, suggesting model selection generalizes across task types.

\begin{table}[t]
\footnotesize
\centering
\caption{
 AD attack reconstruction - steps correctly identified per window. Per-timestamp recall counts and overall average across LLMs. R = reference step count.}
\label{tab:timestamp_recall}
\begin{tabular}{lrrrr}
\toprule
LLMs & 15:30--35 & 15:35--40 & 15:40--46 & Avg Recall\\
 & ($R=11$) & ($R=8$) & ($R=7$) & (\%) \\
\midrule
Claude Sonnet 4 & 11 & 8 & 6 & 96 \\
DeepSeek V3     & 11 & 8 & 6 & 96 \\
Llama 3.1:70b   & 9  & 6 & 6 & 81 \\
Cisco 8B        & 8  & 7 & 6 & 81 \\
GPT-4o          & 5  & 7 & 6 & 69 \\
GPT-4           & 3  & 7 & 6 & 62 \\
GPT-5.2         & 3  & 7 & 5 & 58 \\
GPT-5-mini      & 2  & 6 & 4 & 46 \\
\bottomrule
\end{tabular}

\end{table}

\subsection{Multi-Step Attack Reconstruction}
\label{subsec:cert-abuse-case-study}

\sloppy{We examine how the system reconstructed an ESC1 certificate template abuse attack across five discrete events: certificate enumeration, certificate request with \texttt{SAN:Administrator}, certificate issuance with \texttt{Subject:CN=<user>}, Kerberos TGT as Administrator, and service ticket for \texttt{krbtgt}. Each event was captured by an independent query.}

Given these events as separate chunks, the system produced:

\begin{quote}
\small\itshape
Red enumerated certificates, then requested a certificate with SAN=Administrator. The certificate was issued, allowing impersonation of the Administrator account. Red then obtained a Kerberos TGT as Administrator and requested a service ticket for krbtgt, achieving domain-wide access.
\end{quote}

\sloppy{The exploit mechanism required connecting information across chunks. The certificate \emph{request} chunk contained \texttt{SAN:upn=Administrator}, while the certificate \emph{issued} chunk showed \texttt{Subject:CN=<user>}. The LLM identified this Subject/SAN mismatch as the privilege escalation vector, then linked it to subsequent Kerberos events.}

\finding{4}{The system distinguishes attack from benign activity by recognizing that individually legitimate operations (certificate issuance, Kerberos authentication) constitute exploitation when they enable privilege escalation, demonstrating contextual interpretation beyond pattern matching.}

\subsection{Defense Recommendation Quality}

We assess whether LLM-generated recommendations align with successful defensive actions (\textbf{RQ5}). Table~\ref{tab:ad-defense-windows} summarizes results for Claude Sonnet 4; DeepSeek V3 produces identical recommendations in windows 1--2 and omits RC in window 3.

Two semantic equivalences are applied when scoring. First, recommendations to isolate or quarantine the domain controller are counted as RD, since taking the DC offline achieves the same containment goal. Second, RP is counted as covered when DU is already recommended for the same compromised account: a disabled account cannot authenticate regardless of its password.

\begin{table}[h]
\centering
\caption{Defense Recommendations (Claude Sonnet 4).}
\label{tab:ad-defense-windows}
\footnotesize
\begin{tabular}{lcccc}
\toprule
Window & Blue & LLM & P & R \\
\midrule
15:30--35 & RC, RK, RD        & RC, RK, RD, DU, RP & 3/5 & 3/3 \\
15:35--40 & RK, RD, RP, DU    & DU, RK, RD         & 3/3 & 3/3 \\
15:40--46 & RD, RP, DU, RK    & RC, DU, RD         & 2/3 & 2/3 \\
\bottomrule
\multicolumn{5}{l}{RC=Revoke Cert, RK=Reset KRBTGT, RD=Restart DC,} \\
\multicolumn{5}{l}{DU=Disable User, RP=Reset Password. DU covers RP.}
\end{tabular}
\end{table}

Claude Sonnet 4 achieves 89\% average recall and 76\% average precision without being given the action vocabulary. Recommendations were contextually grounded: certificate revocation after detecting fraudulent certificate issuance, KRBTGT reset after Kerberos ticket abuse, user disable after unauthorized account creation.

\subsection{Operational Feasibility}
\label{subsec:operational-feasibility}

Per-LLM analysis latencies on the AD task are comparable to the per-question latencies reported in \Cref{tab:cost-comparison}, as each window requires a single LLM call; the slowest model (Cisco 8B, under 2 minutes per window) still completes well within the 5-minute window.

This AD scenario represents a fast-moving attack with core exploitation compressed into 60 seconds. The 5-minute window suited this pace, but attack timescales vary widely: ransomware may complete encryption in under 5 minutes, while APT campaigns spread lateral movement over days. Window size must be tuned to threat characteristics. More fundamentally, analysis latency should be measured against response latency rather than attack speed. Defensive actions may require several minutes to execute, and coordinated responses involve human decision-making and approval workflows that typically dominate total incident response time.
The system's sub-minute analysis latency makes detection a small fraction of end-to-end response.

\section{Failure Modes} \label{sec:failure-modes}

Recall misses fall into three disjoint categories.

\emph{Query coverage.}  A reference indicator is not extracted from logs by any feature query.
Fixes are local to the query library: widen filters or increase aggregation depth.
For example, Elasticsearch aggregations return the top-$N$ items per field; on \textit{gozi},
$N{=}10$ silently dropped low-ranked indicators including the 11th C2 IP.
Raising $N$ closes the gap.

\emph{Retrieval and size limits.}
The relevant chunk does not reach the prompt, or the response is truncated.
Fixes tune the retrieval window $k$, the prompt character limit, and the output token budget.
Raising $k$ from 3 to 7 raises Claude's recall on \textit{dec21isc} from 50\% to 100\%; raising the prompt character limit from 20\,K to 60\,K raises \textit{jun21isc} from 27\% to 98\%; raising the output budget from 2\,k to 4\,k tokens removes truncations.

\emph{LLM reasoning.}
The chunk reaches the prompt but the model misinterprets it, fails to integrate across chunks, or formats the output incorrectly.
Fixes are question rewording and model selection.
On \textit{qakbot}, GPT-4 returns the machine account \texttt{desktop-e7fhjs4\$} instead of the user \texttt{damon.bauer};
appending ``(not machine accounts ending in \$)'' to the question resolves the confusion.
On the AD scenario, an enumeration prompt ($``$list each discrete attack step with its timestamp$"$) replaces a narrative prompt and lifts Claude/DeepSeek to 96\% recall.

The first two categories are architectural and affect all models uniformly; the third drives the LLM ranking.

\section{Conclusion}
\label{sec:conclusion}

We presented a retrieval-augmented generation system for security incident analysis that combines targeted query-based filtering with LLM semantic reasoning. Evaluation across eight LLM configurations on malware and Active Directory scenarios demonstrates that RAG preprocessing is essential (no-RAG baselines miss all attack infrastructure), LLM selection determines the performance ceiling once retrieval is properly configured, and DeepSeek V3 achieves near-equivalent accuracy to Claude Sonnet~4 (89\% vs.\ 94\% recall) at 15$\times$ lower cost. Notably, security-specialized models (Cisco Foundation-Sec-8B) do not outperform general-purpose LLMs, suggesting that reasoning capability matters more than domain-specific pretraining for forensic analysis. Our system's key capability is cross-event semantic reasoning: distinguishing coordinated exploitation from individually legitimate operations by correlating evidence across discrete forensic events.

\begin{acks}
This research was funded by the Defense Advanced Research Projects Agency (DARPA), under contract W912CG23C0031.
\end{acks}

\balance

\bibliographystyle{unsrt}
\bibliography{references}

\appendix
\onecolumn

\section*{Appendix: Table of contents}
\startcontents[sections]
\printcontents[sections]{l}{1}{\setcounter{tocdepth}{2}}

\section{Information about the Malware Scenarios} \label{app:malware_info}

File sizes are obtained via \texttt{capinfos}. The number of Security Onion logs is obtained from the Kibana Dashboard associated with Security Onion. Scenarios are sourced from SANS ISC, Palo Alto Unit 42, and Malware-Traffic-Analysis.net.

\begin{table}[ht]
\centering
\caption{Evaluated Malware Scenario Dataset Statistics}
\label{tab:scenario_information}
\small
\begin{tabular}{llrr}
\toprule
\textbf{Dataset} & \textbf{Source} & \textbf{Size} & \textbf{No. of Logs} \\
\midrule
jun21isc       & ISC     & 43.1 MB & 47,991 \\
oct21isc       & ISC     & 46.7 MB & 7,278  \\
apr21isc       & ISC     & 12.9 MB & 5,699  \\
dec21isc       & ISC     & 31.4 MB & 3,869  \\
may21isc       & ISC     & 17.9 MB & 1,544  \\
iced\_id       & Unit-42 & 22.3 MB & 5,701  \\
qakbot         & Unit-42 & 30.5 MB & 4,279  \\
gozi           & Unit-42 & 5.67 MB & 897    \\
redline\_stlr  & Unit-42 & 1.35 MB & 336    \\
lumma\_stealer & MTA     & 25.9 MB & 4,435  \\
fake\_auth     & MTA     & 25.5 MB & 3,694  \\
easy\_123      & MTA     & 6.27 MB & 3,164  \\
warmcookie     & MTA     & 11.4 MB & 2,385  \\
burnincandle   & MTA     & 6.53 MB & 1,114  \\
net\_rat       & MTA     & 20.2 MB & 1,044  \\
koi\_stealer   & MTA     & 2.03 MB & 774    \\
dirty\_rat     & MTA     & 10.9 MB & 744    \\
\midrule
\textbf{Total} & & 321 MB & 94,948 \\
\bottomrule
\multicolumn{4}{l}{\small ISC: SANS ISC;\quad Unit-42: Palo Alto Unit 42;\quad MTA: Malware-Traffic-Analysis.net} \\
\end{tabular}
\end{table}
\section{Prompt Template} \label{app:base_prompt}
\cref{lst:base_prompt} shows the full prompt template sent to the LLM for each forensic question.
The prompt is structured into five components.
First, the agent is assigned the role of a cybersecurity analyst and provided with the specific forensic question to answer.
Second, the agent is instructed to ground its response in specific evidence (IPs, timestamps, hostnames) or explicitly state when the provided data is insufficient.
Third, the agent is required to conclude with a structured output: a \texttt{FINAL ANSWER} field containing the answer, and a \texttt{CITED CHUNKS} field listing the filenames of every retrieved chunk that materially supported the answer, enforcing traceability between the response and the underlying security logs.
Fourth, the \texttt{\{network\_context\}} field supplies the network topology information (\cref{app:network_context}), providing the agent with environmental context such as the LAN segment, domain, and active directory configuration of the monitored network.
Fifth, the Security Data section contains the top-$k$ chunks retrieved from the vector index, each tagged with its source filename in brackets; these are the chunks the agent is expected to cite.

\begin{lstlisting}[style=agentreport, caption={Full prompt template for RAG-based security analysis}, label={lst:base_prompt}]
You are a cybersecurity analyst. Analyze the security data and answer this question:

{question}

Answer with specific evidence (IPs, timestamps, hostnames) or state "Not found in provided data" if insufficient.
Present your analysis and then conclude with EXACTLY two lines, in this order:

FINAL ANSWER = [A, B, C, ...]
CITED CHUNKS = [filename1, filename2, ...]

CITED CHUNKS must list every [filename] block from the Security Data below that
materially supported your FINAL ANSWER. Use the exact filenames shown in
brackets at the start of each chunk (e.g. suricata_alerts_result.json).
Omit chunks you read but did not use. Do not invent filenames.

Network: {network_context}

Security Data:
[chunk_1_filename.json] {chunk_1_content}

[chunk_2_filename.json] {chunk_2_content}

...

[chunk_k_filename.json] {chunk_k_content}

Answer:
\end{lstlisting}







% ...

\section{Sample Network Context} \label{app:network_context}
\cref{lst:ex1_network_context} is an example of network information provided to the LLM to help it understand the network environment.
This information can be found as part of the problem description on the website hosting the PCAP files~\cite{malware_traffic_exercises, unit42_icedid}, and generally includes the LAN, domain, and active directory domain controller.

\begin{lstlisting}[style=agentreport, caption={Network topology information for the Fake Authenticator malware traffic scenario}, label={lst:ex1_network_context}]
- LAN segment range: 10.1.17.0/24 (10.1.17.0 through 10.1.17.255)
- Domain: bluemoontuesday.com
- Active Directory domain controller: 10.1.17.2 - WIN-GSH54QLW48D
- AD environment name: BLUEMOONTUESDAY
- LAN segment gateway: 10.1.17.1
- LAN segment broadcast address: 10.1.17.255
\end{lstlisting}

\section{Questions used for LLM Analysis} \label{app:questions}

Questions are tailored per scenario; the full per-scenario questionnaires are available in the released code.
\cref{lst:questions} shows the core questions asked across all malware scenarios, representative additional questions used in most scenarios, and the enumeration prompt used for the Active Directory attack reconstruction.

\begin{lstlisting}[style=agentreport, caption={Representative forensic questions for malware traffic analysis and the enumeration prompt for Active Directory attack reconstruction}, label={lst:questions}]
# === Core questions (all malware scenarios) ===
What is the IP address of the potentially infected internal host in the LAN?
What is the hostname of the potentially infected machine in the LAN?
What is the Windows human user account name (not machine accounts ending in $) of the potentially infected machine in the LAN?
What are the likely fake or suspicious domains / URLs for initial infection?
What are the suspicious external IP addresses contacted, which might be involved in command-and-control (C2) communication?

# === Additional questions (most malware scenarios) ===
During what hour did the first signs of infection activity appear? Provide the earliest date and time in YYYY-MM-DD HH:MM UTC format.
According to the connection volume statistics, what single external IP address received the most connections from the infected hosts? Is it a C2 server or a legitimate service?
Were any suspicious certificates observed in the encrypted network traffic? If so, identify the domain or IP in the certificate.

# === Active Directory attack reconstruction ===
List each discrete attack step observed in this timeframe, including its timestamp and the specific action taken. Include both new actions and any continuing activity from prior events. For each attack step, recommend a specific defensive action. Enumerate every instance separately, even if the same technique is used multiple times.
\end{lstlisting}

\section{Sample LLM Answers}
\label{app:anthropic_report}

We present sample answers generated by Anthropic on the Fake Authenticator scenario, using a 7-chunks context and the Claude Sonnet 4 model.


\subsection{Identifying the IP of the potentially infected host}
\label{app:q1_anthropic}

\textbf{Question}: What is the IP address of the potentially infected internal host in the LAN?

The LLM Analyzer correctly identified the IP 10.1.17.215 as potentially infected, synthesizing evidence across four distinct data sources. It flagged suspicious domain communications (16 domains contacted solely by this host), high-severity Suricata alerts including Fake Microsoft Teams C2 signatures and PowerShell file requests, Kerberos authentication activity linking the IP to user account \texttt{shutchenson} and hostname \texttt{DESKTOP-L8C5GSJ}, and network context confirming the host's position within the internal LAN segment.

The model concludes with a structured \texttt{FINAL ANSWER = [10.1.17.215]}, providing a machine-parseable answer directly usable for downstream processing. The \texttt{CITED CHUNKS} field lists the four source files that materially supported the answer: \texttt{suspicious\_domains\_result.json}, \texttt{suricata\_alerts\_result.json}, \texttt{kerberos\_hostnames\_result.json}, and \texttt{kerberos\_clients\_result.json}. Of the seven chunks retrieved by the RAG pipeline, the model selectively cited only those directly relevant to the final answer, omitting \texttt{suspicious\_certificates\_result.json} as unused, demonstrating selective, evidence-grounded attribution rather than blanket citation of all retrieved context (\cref{lst:ex1_q1_anthropic}).

\begin{lstlisting}[style=agentreport, caption={Sample LLM answer to the question: What is the infected IP?}, label={lst:ex1_q1_anthropic}]
1. What is the IP address of the potentially infected internal host in the LAN?

Looking at the security data, I need to identify the potentially infected internal host by analyzing the various security indicators.

Analysis:
IP Address 10.1.17.215 shows multiple concerning indicators:

1. Suspicious Domain Communications: This IP is the sole source for connections to 16 different suspicious domains including:
   - "appointedtimeagriculture.com" (suspicious domain name)
   - "authenticatoor.org" (typosquatting-like domain)
   - "google-authenticator.burleson-appliance.net" (suspicious subdomain)
   - "bttrack.com" (potentially BitTorrent related)

2. High-Severity Security Alerts: 10.1.17.215 appears as the top source IP (6 occurrences) in high-severity Suricata alerts, including:
   - "ET MALWARE Fake Microsoft Teams CnC Payload Request (GET)" (2 occurrences)
   - "ET INFO PS1 Powershell File Request" (2 occurrences)
   - "ET INFO TeamViewer Dyngate User-Agent" (2 occurrences)

3. Kerberos Authentication Activity: The IP is associated with:
   - User account "shutchenson" (11 authentication events)
   - Computer accounts "DESKTOP-L8C5GSJ$" and "desktop-l8c5gsj$" (14 total events)
   - Hostname "desktop-l8c5gsj"

4. Network Context: 10.1.17.215 is within the internal LAN segment (10.1.17.0/24) and shows extensive outbound communication patterns consistent with potential malware behavior.

The combination of malware-related alerts, suspicious domain communications, PowerShell file requests, and potential C&C communications strongly indicates this host is compromised.

FINAL ANSWER = [10.1.17.215]

CITED CHUNKS = [suspicious_domains_result.json, suricata_alerts_result.json, kerberos_hostnames_result.json, kerberos_clients_result.json]

RETRIEVED CHUNKS = [suspicious_domains_result.json, suspicious_certificates_result.json, kerberos_hostnames_result.json, suspicious_certificates_result.json, suricata_alerts_result.json, suricata_alerts_result.json, kerberos_clients_result.json]
\end{lstlisting}


\subsection{Identifying C2 infrastructure}
\label{app:example_c2_question}

\textbf{Question:} What are the suspicious external IP addresses contacted, which might be involved in command-and-control (C2) communication?

The LLM Analyzer correctly identified four suspicious external IPs involved in C2 communication: 5.252.153.241, 185.188.32.26, 45.125.66.32, and 45.125.66.252. The model applied a confidence-tiered reasoning approach, distinguishing high-confidence C2 IPs confirmed by Spamhaus DROP blocklist hits and high-severity Suricata alerts from medium-confidence IPs flagged by suspicious certificate characteristics (IP-based and self-signed certificates). This tiered analysis demonstrates evidence-weighted reasoning, prioritizing corroborated indicators over individually suspicious signals. The \texttt{CITED CHUNKS} field correctly attributes the answer to two source files: \texttt{suspicious\_certificates\_result.json} and \texttt{suricata\_alerts\_result.json}, selectively omitting the remaining five retrieved chunks as unused (\cref{lst:ex1_q5_anthropic}).

\begin{lstlisting}[style=agentreport, caption={Sample LLM Answer to the question: What is the C2?}, label={lst:ex1_q5_anthropic}]
5. What are the suspicious external IP addresses contacted, which might be involved in command-and-control (C2) communication?

Looking at the security data, I can identify several suspicious external IP addresses that are likely involved in command-and-control (C2) communication based on multiple indicators:

Analysis

High-Confidence C2 IPs:
1. 5.252.153.241 - This IP appears 4 times in high-severity alerts and is flagged by "ET DROP Spamhaus DROP Listed Traffic Inbound group 5", indicating it's on a known malicious IP blocklist.
2. 185.188.32.26 - This IP appears 2 times in high-severity alerts and is also flagged by the same Spamhaus DROP alert, confirming malicious activity.

Medium-Confidence Suspicious IPs:
3. 45.125.66.32 and 45.125.66.252 - These IPs are using suspicious IP-based certificates (rather than domain-based certificates) and self-signed certificates, which are common C2 infrastructure characteristics. They also appear as sources in high-severity alerts.

Additional Context:
- The alerts include "ET MALWARE Fake Microsoft Teams CnC Payload Request (GET)" which directly indicates C2 communication attempts
- "ET INFO PS1 Powershell File Request" suggests potential malicious PowerShell activity
- All suspicious activity appears to involve internal host 10.1.17.215 (desktop-l8c5gsj) as either source or destination

The combination of Spamhaus blocklist hits, malware-specific signatures, and suspicious certificate usage provides strong evidence these IPs are involved in C2 operations.

FINAL ANSWER = [5.252.153.241, 185.188.32.26, 45.125.66.32, 45.125.66.252]

CITED CHUNKS = [suspicious_certificates_result.json, suricata_alerts_result.json]

RETRIEVED CHUNKS = [suspicious_domains_result.json, suspicious_certificates_result.json, suspicious_certificates_result.json, suricata_alerts_result.json, suspicious_certificates_result.json, suricata_alerts_result.json, kerberos_hostnames_result.json]
\end{lstlisting}


\section{Aggregation Queries}
\label{appendix:aggregation-queries}

The following Elasticsearch queries produce the aggregated results shown in Section~\ref{sec:example}. Each query pairs a filter with aggregations that compress matching events into ranked summaries.

\subsection{Kerberos Client Aggregation}

This query extracts usernames from Kerberos authentication events (port 88) and groups them by source IP address.

\begin{lstlisting}[style=agentreport, caption={Kerberos Client Query}, label={lst:kerberos_client_query}]
{
  "query": {
    "bool": {
      "must": [
        {"term": {"destination.port": 88}},
        {"range": {"@timestamp": {"gte": "...", "lte": "..."}}}
      ]
    }
  },
  "aggs": {
    "client_accounts": {
      "terms": {
        "script": {
          "source": "def msg = doc['message.keyword'].value;
                     def pattern = /\"client\":\"([^\\/\"]+)/;
                     def matcher = pattern.matcher(msg);
                     if (matcher.find()) return matcher.group(1);"
        },
        "size": 20
      },
      "aggs": {
        "source_ips": {
          "terms": {"field": "source.ip", "size": 10}
        }
      }
    }
  },
  "size": 0
}
\end{lstlisting}

\subsection{Suricata High-Severity Alert Aggregation}

This query filters for IDS alerts from internal hosts and aggregates by signature name, source IP, and destination IP, with a nested filter for high-severity events.

\begin{lstlisting}[style=agentreport, caption={Suricata Alerts Query}, label={lst:suricata_alerts_query}]
{
  "query": {
    "bool": {
      "must": [
        {"bool": {"should": [
          {"term": {"event.dataset": "suricata.alert"}},
          {"term": {"event.module": "suricata"}}
        ]}},
        {"range": {"source.ip": {"gte": "10.0.0.0", "lte": "172.31.255.255"}}}
      ]
    }
  },
  "aggs": {
    "high_severity_alerts": {
      "filter": {"range": {"rule.severity": {"gte": 1, "lte": 2}}},
      "aggs": {
        "high_severity_signatures": {
          "terms": {"field": "rule.name", "size": 10}
        },
        "high_severity_sources": {
          "terms": {"field": "source.ip", "size": 10}
        },
        "high_severity_destinations": {
          "terms": {"field": "destination.ip", "size": 5}
        }
      }
    }
  },
  "size": 0
}
\end{lstlisting}

\subsection{File Download Aggregation}

This query identifies file downloads by filtering for common download ports and file extensions, then uses a script to reconstruct HTTP request patterns from log messages.

\begin{lstlisting}[style=agentreport, caption={File Download Query}, label={lst:file_download_query}]
{
  "query": {
    "bool": {
      "must": [
        {"terms": {"destination.port": [80, 443, 8080, 8443]}}
      ],
      "should": [
        {"wildcard": {"message": "*.ps1*"}},
        {"wildcard": {"message": "*.exe*"}},
        {"wildcard": {"http.uri": "*.ps1*"}},
        {"wildcard": {"http.uri": "*.exe*"}}
      ]
    }
  },
  "aggs": {
    "http_requests": {
      "terms": {
        "script": {
          "source": "def msg = doc['message.keyword'].value;
                     def pattern = /GET[ ]+([^\\s]+)/;
                     def matcher = pattern.matcher(msg);
                     if (matcher.find()) {
                       return doc['source.ip'].value + ' -> ' +
                              doc['destination.ip'].value + ':' +
                              doc['destination.port'].value + ' : ' +
                              matcher.group(1);
                     }
                     return 'no_http_uri';"
        }
      }
    },
    "file_types": {
      "terms": {
        "script": {
          "source": "def uri = doc['http.uri'].value;
                     if (uri.contains('.exe')) return 'exe';
                     if (uri.contains('.ps1')) return 'powershell';
                     if (uri.contains('.msi')) return 'msi';
                     return 'other';"
        }
      }
    }
  },
  "size": 0
}
\end{lstlisting}

\subsection{LLM Analysis Output}

When the aggregation results are retrieved and provided to the LLM, it corroborates across sources to answer forensic questions. The following output was generated by Claude Sonnet 4 (7-chunk context) for the question: \textit{``What is the Windows user account name of the potentially infected machine in the LAN?''}

The LLM Analyzer correctly identified \texttt{shutchenson} as the Windows user account of the potentially infected machine. The model followed a structured three-step reasoning chain: first identifying the compromised IP (10.1.17.215) from Suricata alert patterns, then correlating it to hostname \texttt{desktop-l8c5gsj} via Kerberos hostname data, and finally disambiguating the human user account from machine accounts by explicitly filtering out accounts ending in \texttt{\$}. The model cited three source files in \texttt{CITED CHUNKS}, selectively omitting four of the seven retrieved chunks as unused, demonstrating focused attribution across the cross-source reasoning chain (\cref{lst:corroboration_answer}).

\begin{lstlisting}[style=agentreport, caption={LLM analysis demonstrating cross-query corroboration}, label={lst:corroboration_answer}]
3. What is the Windows human user account name (not machine accounts ending in $) of the potentially infected machine in the LAN?

Looking at the security data, I need to identify the Windows human user account name of the potentially infected machine.

Analysis

Potentially Infected Machine Identification:
From the Suricata alerts, I can see several concerning security alerts including:
- "ET MALWARE Fake Microsoft Teams CnC Payload Request (GET)" (2 occurrences)
- "ET DROP Spamhaus DROP Listed Traffic Inbound group 5" (2 occurrences)
- "ET INFO PS1 Powershell File Request" (2 occurrences)

Machine and User Correlation:
The Kerberos data shows all authentication activity is coming from IP 10.1.17.215, which corresponds to hostname "desktop-l8c5gsj" based on the hostname aggregation data.

User Account Analysis:
From the Kerberos client accounts data for IP 10.1.17.215, I can see three accounts:
1. "shutchenson" - 11 authentication events (human user account)
2. "DESKTOP-L8C5GSJ$" - 10 authentication events (machine account - ends in $)
3. "desktop-l8c5gsj$" - 4 authentication events (machine account - ends in $)

Suspicious Activity Correlation:
The suspicious domains data shows that IP 10.1.17.215 is accessing multiple suspicious domains, and this same IP is generating malware alerts in Suricata. The machine accounts ending in "$" are excluded per the question requirements, leaving "shutchenson" as the human user account.

Evidence Summary:
- Infected machine IP: 10.1.17.215
- Hostname: desktop-l8c5gsj
- Malware alerts: "ET MALWARE Fake Microsoft Teams CnC Payload Request"
- Human user account: shutchenson (11 Kerberos authentications from 10.1.17.215)

FINAL ANSWER = shutchenson

CITED CHUNKS = [kerberos_clients_result.json, suricata_alerts_result.json, kerberos_hostnames_result.json]

RETRIEVED CHUNKS = [kerberos_hostnames_result.json, suricata_alerts_result.json, suspicious_domains_result.json, kerberos_clients_result.json, suspicious_certificates_result.json, kerberos_full_services_result.json, suspicious_certificates_result.json]
\end{lstlisting}

\section{Comparative Analysis of LLM Reasoning}
\label{appendix:llm-reasoning}

To illustrate how different LLMs approach security incident analysis, we compare outputs from three providers on the same question from the Koi Stealer scenario: \textit{``What are the suspicious external IP addresses contacted, which might be involved in command-and-control (C2) communication?''} This question requires synthesizing evidence from multiple data sources (Suricata alerts, DNS logs, connection data) and distinguishing confirmed C2 infrastructure from merely suspicious traffic.

The reference answer for this scenario is \texttt{79.124.78.197} as the C2 server, which received 49 high-severity alerts including explicit Koi Stealer signatures.


\subsection{DeepSeek V3: Precision-Focused Analysis}

DeepSeek produces a structured analysis that systematically evaluates evidence strength and explicitly rejects weak indicators. Faced with multiple external IPs in the retrieved context, DeepSeek identifies 79.124.78.197 as the sole confirmed C2 server based on two convergent signals: 49 high-severity Suricata alerts with explicit Koi Stealer C2 signatures, and 218 connections in the high-activity external destination data. It explicitly considers other external IPs contacted by the victim host but rejects them individually, noting the absence of corroborating alert or signature data and even identifying benign infrastructure (Microsoft and Azure IPs) among the candidates. This demonstrates evidence-weighted reasoning: the model distinguishes between ``contacted external IPs'' and ``confirmed C2 infrastructure,'' prioritizing corroborated indicators over volume, and arrives at a single, high-confidence answer.

\begin{lstlisting}[style=agentreport, caption={DeepSeek V3 analysis of C2 indicators}, label={lst:deepseek-c2}]
5. What are the suspicious external IP addresses contacted, which might be involved in command-and-control (C2) communication?

**Analysis**

The question asks for suspicious external IP addresses that might be involved in command-and-control (C2) communication. The provided data shows network activity primarily from the internal host 172.17.0.99 (hostname desktop-rnvo9at, user afletcher).

Key evidence for identifying suspicious external IPs:

1. **Suricata Alerts:** The high_severity_alerts aggregation shows 49 alerts for the signature "ET HUNTING GENERIC SUSPICIOUS POST to Dotted Quad with Fake Browser 1". The destination for these alerts is the external IP 79.124.78.197. Furthermore, there are single alerts for "ET ATTACK_RESPONSE Koi Loader/Stealer CnC Config Inbound" and "ET MALWARE Win32/Koi Stealer CnC Checkin (GET)", which are explicitly named as C2-related. The source/destination data for these high-severity alerts points to communication between 172.17.0.99 and 79.124.78.197.

2. **High Activity Logs:** The high_activity_external_dest_result.json data strongly supports this, showing that internal host 172.17.0.99 contacted 35 unique external destinations. Its top destination by a massive margin (218 connections) is 79.124.78.197. This high volume of traffic to a single external IP, coupled with the C2-themed Suricata alerts to/from that same IP, is highly indicative of C2 communication.

3. **Other External IPs:** The same high-activity data lists other external IPs contacted by 172.17.0.99 (23.45.119.144, 204.79.197.203, 52.113.194.132, 23.195.212.189). However, there is no direct alert or signature data linking these IPs to malicious activity. Some may be benign (e.g., 204.79.197.203 is a Microsoft IP, 52.113.194.132 is likely Azure). Without additional context like suspicious domains or alerts tied to them, they do not meet the threshold for "suspicious... involved in C2" based on the provided data.

**Conclusion:** The only external IP with direct, multiple pieces of evidence linking it to potential C2 activity is 79.124.78.197.

FINAL ANSWER = [79.124.78.197]

CITED CHUNKS = [suricata_alerts_result.json, high_activity_external_dest_result.json]

RETRIEVED CHUNKS = [suspicious_domains_result.json, suricata_alerts_result.json, suricata_alerts_result.json, kerberos_hostnames_result.json, kerberos_full_services_result.json, kerberos_clients_result.json, high_activity_external_dest_result.json]
\end{lstlisting}


\subsection{Cisco Foundation-Sec-8B: Verbose Chain-of-Thought}

The security-specialized Cisco model produces verbose chain-of-thought reasoning that explicitly works through each data source before synthesizing a conclusion. Like DeepSeek, it correctly identifies \texttt{79.124.78.197} as the sole C2 server, but through a longer reasoning path that narrates its inspection of each retrieved chunk in sequence.

\begin{lstlisting}[style=agentreport, caption={Cisco Foundation-Sec-8B analysis of C2 indicators (abbreviated)}, label={lst:cisco-c2}]
The user is asking me to identify suspicious external IP addresses that might
be involved in Command-and-Control (C2) communication based on the provided
security data. I need to analyze the given JSON files and extract relevant
information.

First, I'll examine the `suricata_alerts_result.json` file under the
high_severity_alerts aggregation... Within this, there's a high_severity_sources
bucket. Looking inside it, I see an entry for `79.124.78.197` with a doc_count
of 49. The key associated with this IP is "ET HUNTING GENERIC SUSPICIOUS POST
to Dotted Quad with Fake Browser 1". This signature suggests potential malicious
activity, possibly related to C2...

[...sequential analysis of kerberos, high-activity destination, and domain data...]

Considering all the evidence, `79.124.78.197` stands out due to its association
with a high-severity Suricata alert signature. No other IP in the data provides
such clear and direct evidence of involvement in C2 communication.

Based on the provided security data, the following suspicious external IP
addresses were identified as potentially involved in C2 communication:

Suspicious External IP Addresses:
79.124.78.197

**FINAL ANSWER = [79.124.78.197]**

**CITED CHUNKS = [suricata_alerts_result.json]**

RETRIEVED CHUNKS = [suspicious_domains_result.json, suricata_alerts_result.json, suricata_alerts_result.json, kerberos_hostnames_result.json, kerberos_full_services_result.json, kerberos_clients_result.json, high_activity_external_dest_result.json]
\end{lstlisting}

Cisco arrives at the same correct answer as DeepSeek, but its reasoning is substantially more verbose: it narrates its inspection of each chunk sequentially rather than synthesizing across sources. The response cites only one chunk (\texttt{suricata\_alerts\_result.json}), whereas DeepSeek cites two, suggesting Cisco anchors on the strongest single signal rather than cross-validating. The output formatting is already atypical here: Cisco wraps the structured fields in Markdown bold (\texttt{**...**}) rather than emitting plain text. Across the full evaluation, Cisco's formatting varied further---sometimes omitting brackets around answer lists, sometimes dropping \texttt{CITED CHUNKS} entirely---requiring manual review during scoring.


\subsection{GPT-5.2: Enumerative Listing}

GPT-5.2 produces a concise response that lists external IPs without differentiating evidence strength or justifying individual inclusions. The answer correctly identifies 79.124.78.197 but bundles it alongside four other IPs drawn from the high-activity destination data, including 204.79.197.203 and 52.113.194.132 which are Microsoft and Azure infrastructure respectively. No reasoning is provided for why these IPs are considered suspicious, nor does the model acknowledge the benign nature of known cloud provider addresses. Unlike DeepSeek, which explicitly evaluated and rejected these same IPs, GPT-5.2 treats the question as a data extraction task, listing all high-activity destinations without evaluating their relevance to the specific question of C2 communication.

\begin{lstlisting}[style=agentreport, caption={GPT-5.2 analysis of C2 indicators}, label={lst:gpt52-c2}]
5. What are the suspicious external IP addresses contacted, which might be involved in command-and-control (C2) communication?

The suspicious external IP addresses contacted, which might be involved in command-and-control (C2) communication, are:
- 79.124.78.197
- 23.45.119.144
- 204.79.197.203
- 52.113.194.132
- 23.195.212.189

FINAL ANSWER = [79.124.78.197, 23.45.119.144, 204.79.197.203, 52.113.194.132, 23.195.212.189]

CITED CHUNKS = [suricata_alerts_result.json, high_activity_external_dest_result.json]

RETRIEVED CHUNKS = [suspicious_domains_result.json, suricata_alerts_result.json, suricata_alerts_result.json, kerberos_hostnames_result.json, kerberos_full_services_result.json, kerberos_clients_result.json, high_activity_external_dest_result.json]
\end{lstlisting}

GPT-5.2 conflates high-activity destination IPs with confirmed C2 infrastructure, reporting all external destinations above an activity threshold without evaluating their individual threat relevance. Including known Microsoft and Azure IPs (204.79.197.203, 52.113.194.132) as suspicious C2 candidates represents a failure to apply basic threat contextualization. The model treats the question as a data extraction task rather than an analytical one, reporting IPs present in the retrieved chunks without evaluating their relevance to the specific question of C2 communication.

\section{Precision-Recall Analysis}
\label{appendix:precision}

This appendix provides precision-recall metrics for set-valued questions across all 17 malware scenarios and 8 LLM providers. Set-valued questions are those whose reference answer contains more than one indicator, specifically questions about suspicious domains and URLs for initial infection, and suspicious external IPs for C2 communication.

\begin{itemize}
    \item \textbf{Qs} denotes the number of set-valued questions in that scenario;
    \item \textbf{Ref} denotes the total number of reference indicators across those questions;
    \item \textbf{N} denotes the total number of indicators predicted by the model;
    \item \textbf{TP/N} shows precision as a ratio of true positives to predicted indicators;
    \item \textbf{Rec.} denotes mean recall averaged across set-valued questions within each scenario.
\end{itemize}

In cases where the model answers in sentence form rather than a delimiter-separated list (e.g., ``45.131.214.85 is a C2 server''), \textbf{N} is adjusted to reflect the number of matched reference tokens rather than the number of parsed predicted items. The reference set captures \emph{confirmed} malicious indicators from published solutions; several indicators reported as false positives exhibit genuinely suspicious characteristics, so true precision may be higher than reported.

The \textbf{Avg.} row reports aggregate precision as total TP / total N across all scenarios, and macro-averaged recall (mean of per-scenario recall values, consistent with the main evaluation).

Claude and DeepSeek achieve the highest aggregate precision (107/141 and 90/120 respectively) alongside strong recall, reflecting focused, evidence-grounded responses. Cisco and GPT models show substantially lower precision, driven by heavy over-prediction on scenarios such as \texttt{fake\_auth} (Cisco: 6/51) and \texttt{jun21isc} (GPT models: 11--12/72).

\begin{table}[htbp]
\centering
\caption{Precision-Recall Analysis: Set-Valued Questions Across All Scenarios (single-answer questions excluded)}
\label{tab:precision-recall-full}
\scriptsize
\setlength{\tabcolsep}{3pt}
\begin{tabular}{lrr|rrr|rrr|rrr|rrr|rrr|rrr|rrr|rrr}
\toprule
 & & & \multicolumn{3}{c|}{\textbf{Claude}} & \multicolumn{3}{c|}{\textbf{DeepSeek}} & \multicolumn{3}{c|}{\textbf{Ollama}} & \multicolumn{3}{c|}{\textbf{Cisco}} & \multicolumn{3}{c|}{\textbf{GPT-4}} & \multicolumn{3}{c|}{\textbf{GPT-4o}} & \multicolumn{3}{c|}{\textbf{GPT-5-mini}} & \multicolumn{3}{c}{\textbf{GPT-5.2}} \\
\cmidrule(lr){4-6} \cmidrule(lr){7-9} \cmidrule(lr){10-12} \cmidrule(lr){13-15} \cmidrule(lr){16-18} \cmidrule(lr){19-21} \cmidrule(lr){22-24} \cmidrule(lr){25-27}
\textbf{Dataset} & \textbf{Qs} & \textbf{Ref} & \textbf{N} & \textbf{TP/N} & \textbf{Rec.} & \textbf{N} & \textbf{TP/N} & \textbf{Rec.} & \textbf{N} & \textbf{TP/N} & \textbf{Rec.} & \textbf{N} & \textbf{TP/N} & \textbf{Rec.} & \textbf{N} & \textbf{TP/N} & \textbf{Rec.} & \textbf{N} & \textbf{TP/N} & \textbf{Rec.} & \textbf{N} & \textbf{TP/N} & \textbf{Rec.} & \textbf{N} & \textbf{TP/N} & \textbf{Rec.} \\
\midrule
apr21isc      & 2 & 8  & 6  & 4/6   & 66.5  & 5  & 4/5   & 66.5  & 6  & 4/6   & 66.5  & 4  & 2/4   & 33.5  & 7  & 3/7   & 41.5  & 7  & 3/7   & 41.5  & 7  & 3/7   & 41.5  & 7  & 3/7   & 41.5  \\
burnincandle  & 3 & 14 & 16 & 14/16 & 100.0 & 14 & 11/14 & 83.3  & 12 & 10/12 & 62.7  & 7  & 5/7   & 45.7  & 10 & 10/10 & 62.7  & 10 & 10/10 & 62.7  & 10 & 10/10 & 62.7  & 10 & 10/10 & 62.7  \\
dec21isc      & 3 & 6  & 10 & 6/10  & 100.0 & 10 & 6/10  & 100.0 & 9  & 5/9   & 83.3  & 7  & 5/7   & 83.3  & 8  & 4/8   & 66.7  & 8  & 4/8   & 66.7  & 8  & 4/8   & 66.7  & 8  & 4/8   & 66.7  \\
dirty\_rat    & 2 & 6  & 4  & 3/4   & 62.5  & 5  & 1/5   & 12.5  & 3  & 0/3   & 0.0   & 4  & 0/4   & 0.0   & 11 & 1/11  & 25.0  & 11 & 1/11  & 25.0  & 11 & 1/11  & 25.0  & 11 & 1/11  & 25.0  \\
easy\_123     & 1 & 2  & 2  & 2/2   & 100.0 & 2  & 2/2   & 100.0 & 1  & 1/1   & 50.0  & 2  & 2/2   & 100.0 & 1  & 1/1   & 50.0  & 1  & 1/1   & 50.0  & 1  & 1/1   & 50.0  & 1  & 1/1   & 50.0  \\
fake\_auth    & 3 & 7  & 8  & 7/8   & 100.0 & 8  & 7/8   & 100.0 & 7  & 6/7   & 83.3  & 51 & 6/51  & 83.3  & 13 & 3/13  & 50.0  & 13 & 3/13  & 50.0  & 6  & 3/6   & 50.0  & 13 & 3/13  & 50.0  \\
gozi          & 3 & 17 & 9  & 9/9   & 65.0  & 11 & 9/11  & 65.0  & 10 & 8/10  & 48.3  & 9  & 7/9   & 59.0  & 8  & 8/8   & 48.3  & 8  & 6/8   & 42.3  & 8  & 6/8   & 42.3  & 8  & 6/8   & 42.3  \\
iced\_id      & 2 & 6  & 7  & 4/7   & 75.0  & 6  & 3/6   & 62.5  & 5  & 3/5   & 50.0  & 5  & 3/5   & 50.0  & 32 & 3/32  & 50.0  & 32 & 3/32  & 50.0  & 32 & 3/32  & 50.0  & 32 & 3/32  & 50.0  \\
jun21isc      & 6 & 21 & 29 & 20/29 & 97.2  & 16 & 14/16 & 73.7  & 22 & 12/22 & 59.8  & 12 & 6/12  & 34.7  & 72 & 11/72 & 55.5  & 72 & 12/72 & 63.8  & 73 & 11/73 & 55.5  & 72 & 11/72 & 55.5  \\
koi\_stealer  & 1 & 2  & 2  & 2/2   & 100.0 & 2  & 2/2   & 100.0 & 1  & 1/1   & 50.0  & 2  & 2/2   & 100.0 & 1  & 1/1   & 50.0  & 1  & 1/1   & 50.0  & 1  & 1/1   & 50.0  & 1  & 1/1   & 50.0  \\
lumma\_stealer & 2 & 4 & 6  & 4/6   & 100.0 & 7  & 4/7   & 100.0 & 4  & 3/4   & 75.0  & 3  & 2/3   & 50.0  & 3  & 2/3   & 50.0  & 3  & 2/3   & 50.0  & 3  & 2/3   & 50.0  & 3  & 2/3   & 50.0  \\
may21isc      & 1 & 2  & 2  & 2/2   & 100.0 & 2  & 2/2   & 100.0 & 1  & 1/1   & 50.0  & 2  & 1/2   & 50.0  & 1  & 1/1   & 50.0  & 1  & 1/1   & 50.0  & 1  & 1/1   & 50.0  & 1  & 1/1   & 50.0  \\
net\_rat      & 2 & 4  & 8  & 4/8   & 100.0 & 4  & 3/4   & 75.0  & 3  & 2/3   & 50.0  & 4  & 0/4   & 0.0   & 6  & 3/6   & 75.0  & 5  & 2/5   & 50.0  & 1  & 0/1   & 0.0   & 5  & 2/5   & 50.0  \\
oct21isc      & 5 & 15 & 15 & 15/15 & 100.0 & 12 & 12/12 & 85.0  & 12 & 11/12 & 75.0  & 38 & 5/38  & 36.6  & 28 & 8/28  & 53.4  & 16 & 12/16 & 73.4  & 28 & 8/28  & 53.4  & 16 & 12/16 & 73.4  \\
qakbot        & 2 & 7  & 6  & 3/6   & 60.0  & 4  & 3/4   & 60.0  & 5  & 2/5   & 35.0  & 11 & 1/11  & 25.0  & 16 & 1/16  & 25.0  & 7  & 0/7   & 0.0   & 16 & 1/16  & 25.0  & 16 & 0/16  & 0.0   \\
redline\_stlr & 2 & 4  & 4  & 3/4   & 75.0  & 7  & 3/7   & 75.0  & 4  & 1/4   & 25.0  & 3  & 1/3   & 25.0  & 6  & 3/6   & 75.0  & 6  & 3/6   & 75.0  & 4  & 2/4   & 50.0  & 6  & 3/6   & 75.0  \\
warmcookie    & 2 & 6  & 7  & 5/7   & 87.5  & 5  & 4/5   & 75.0  & 4  & 1/4   & 25.0  & 5  & 2/5   & 50.0  & 11 & 2/11  & 37.5  & 11 & 2/11  & 37.5  & 11 & 2/11  & 37.5  & 11 & 2/11  & 37.5  \\
\midrule
\textbf{Avg.} & & & \multicolumn{1}{r}{---} & 107/141 & 87.6 & \multicolumn{1}{r}{---} & 90/120 & 78.4 & \multicolumn{1}{r}{---} & 71/109 & 52.3 & \multicolumn{1}{r}{---} & 50/169 & 48.6 & \multicolumn{1}{r}{---} & 65/234 & 50.9 & \multicolumn{1}{r}{---} & 66/212 & 49.3 & \multicolumn{1}{r}{---} & 59/221 & 44.7 & \multicolumn{1}{r}{---} & 65/221 & 48.8 \\
\bottomrule
\end{tabular}
\end{table}

\section{IcedID Dataset Complexity}
\label{appendix:icedid}
Here we present some aspects that make the IcedID malware particularly challenging, and how our system uncovers attack indicators.

\paragraph{\textbf{Multi-Host Environment with Mixed Traffic.}}
The capture contains traffic from three Windows hosts in an Active Directory environment, with only one confirmed compromise.
DESKTOP-SFF9LJF (10.4.19.136) follows the documented IcedID chain: Firebase-hosted installer, HTTP binary retrieval, persistent HTTPS C2 beaconing, and BackConnect activity.
Other hosts in the environment (including DESKTOP-RETP4BU at 10.4.19.138) generate background traffic that includes DNS queries and web requests unrelated to the IcedID infection.
This multi-host traffic creates analytical noise, as there is no structural boundary separating the infected host's footprint from routine AD activity.

\paragraph{\textbf{Volume as a Misleading Signal.}}
Domain join sequences and routine AD operations from uninfected hosts produce heavy Kerberos and LDAP bursts that are volumetrically louder than the infected host's quieter but diagnostically significant IcedID C2 beaconing, making raw traffic volume an unreliable primary indicator in this capture.

\paragraph{\textbf{Reference Set Scope.}}
The Unit 42 reference answers cover six IOCs for a single host.
During evaluation, our system surfaced additional indicators confirmed valid upon manual investigation:
\begin{itemize}
    \item \textbf{Additional C2 domain.} Our system cited \texttt{spakernakurs.com} resolving to \texttt{104.168.53.18}, identical to \texttt{askamoshopsi.com}---and absent from the Unit 42 writeup. Manual lookup confirmed the domain resolves to shared IcedID infrastructure, consistent with a rotation or replacement C2 domain.
    \item \textbf{Beaconing regularity.} Our system's classification of \texttt{skansnekssky.com} as persistent C2 activity prompted tshark verification, confirming ${\sim}$302-second intervals with sub-second jitter across 69+ sessions, a pattern unquantified in the reference answers.
\end{itemize}

\paragraph{\textbf{Query Adaptation.}}
The IcedID installer delivery via Firebase Storage (\texttt{firebasestorage.googleapis.com}) required query modification: the standard \texttt{*googleapis.com*} exclusion for noise reduction would suppress the entire infection vector. 
We replaced broad wildcards with surgical per-subdomain exclusions and added two aggregations: (1)~direct-IP TLS connections, recovering BackConnect C2 at 193.149.176.100, and (2)~HTTP requests to raw IP hosts, recovering the redirect server at 80.77.25.175.
This adaptation illustrates how attackers exploit common filtering heuristics and the need for dataset-aware query refinement.

\section{MITRE ATT\&CK Technique Coverage}
\label{app:mitre-coverage}

The two query libraries cover 16 of $\sim$200 MITRE ATT\&CK Enterprise techniques. Coverage is scoped by data source and attack type: our current sources are network PCAPs (via Zeek/Suricata) and Windows Event Logs.
Broader coverage requires adding data sources and corresponding queries, and our RAG pipeline can be easily extended with additional sources and indicators. For example, endpoint EDR would cover process-level techniques (T1055 Process Injection, T1003 Credential Dumping), email gateway logs would cover initial delivery (T1566 Phishing), and proxy logs would cover web-based exfiltration (T1567 Exfiltration Over Web Service), etc.
The malware-traffic library covers 6 techniques: T1071 C2 beaconing, T1573 Encrypted Channel, T1105 file downloads, T1568 suspicious domains, T1078 Kerberos auth, T1041 high-volume connections.
Table~\ref{tab:mitre-mapping-compact} shows the AD attack query library with its MITRE ATT\&CK mappings.

\begin{table}[htbp]
\centering
\caption{Query Library for Active Directory Attack Detection}
\label{tab:mitre-mapping-compact}
\footnotesize
\begin{tabular}{@{}ll@{}}
\toprule
\textbf{Query} & \textbf{MITRE ATT\&CK} \\
\midrule
PowerShell Injection           & T1059 -- Command and Scripting Interpreter \\
File Drop                      & T1204 -- User Execution \\
Certificate Enumeration        & T1649 -- Steal/Forge Authentication Certificates \\
Certificate Request            & T1649 -- Steal/Forge Authentication Certificates \\
Certificate Issued             & T1649 -- Steal/Forge Authentication Certificates \\
Kerberos Authentication        & T1558 -- Steal/Forge Kerberos Tickets \\
Kerberos Service Tickets       & T1558 -- Steal/Forge Kerberos Tickets \\
Service Installation           & T1543 -- Create/Modify System Process \\
User Account Creation          & T1136 -- Create Account \\
Security Group Modification    & T1098 -- Account Manipulation \\
SSH to SCADA Server            & T1021.004 -- Remote Services: SSH \\
SCADA Service Stopped          & T1489 -- Service Stop \\
\bottomrule
\end{tabular}
\end{table}

\section{Choice of Analysis Window}
\label{app:window-justification}

The attack spans $\sim$15 minutes; we split it into three 5-minute windows.
A 5-minute window produces $\sim$40\,KB of extracted data, which fits in the prompt. The full 15-minute window produces $\sim$100\,KB, exceeding the prompt limit: Claude drops from 96\% to 62\%, with missed steps due to both data truncation and the model summarizing rather than enumerating steps in a longer context. Shorter windows (1--2 min) would fragment continuing activity across LLM calls.

\end{document}